\documentclass[useAMS,usenatbib]{mn2e}
\usepackage{graphicx}

\def\rmsLz{\left\langle(\Delta L_z)^2\right\rangle^{1/2}}

\def\etal{{\it  et al.}}
\def\eg{{\it e.g.}}
\def\ie{{\it i.e.}}
\def\DF{{\small DF}}
\long\def\Ignore#1{\relax}

\title[Radial Migration in Thick Discs]
  {Radial Migration in Galactic Thick Discs}
\author[M. Solway, J. A. Sellwood \& R. Sch\"onrich]
  {Michael Solway,$^1$
   J. A. Sellwood,$^1$
   Ralph Sch\"onrich,$^2$ \\
  $^1$Rutgers University, Department of Physics \& Astronomy,
      136 Frelinghuysen Road, Piscataway, NJ 08854-8019, US \\
  $^2$Max-Planck-Institut f\"ur Astrophysik,
      Karl-Schwarzschild-Str. 1, 85741 Garching, Germany}

\pagerange{\pageref{firstpage}--\pageref{lastpage}} \pubyear{2011}

\def\LaTeX{L\kern-.36em\raise.3ex\hbox{a}\kern-.15em
    T\kern-.1667em\lower.7ex\hbox{E}\kern-.125emX}

\begin{document}

\label{firstpage}

\maketitle

\begin{abstract}
We present a study of the extent to which the Sellwood \& Binney
radial migration of stars is affected by their vertical motion about
the midplane.  We use both controlled simulations in which only a
single spiral mode is excited, as well as slightly more realistic
cases with multiple spiral patterns and a bar.  We find that rms
angular momentum changes are reduced by vertical motion, but rather
gradually, and the maximum changes are almost as large for thick disc
stars as for those in a thin disc.  We find that particles in
simulations in which a bar forms suffer slightly larger angular
momentum changes than in comparable cases with no bar, but the
cumulative effect of multiple spiral events still dominates.  We have
determined that vertical action, and not vertical energy, is conserved
on average during radial migration.
\end{abstract}


\section{Introduction}
It has recently become clear that galaxy discs evolve significantly
over time due to both internal and external influences \citep[see][for
  reviews]{V11,S10}.  Scattering by giant molecular clouds
\citep[\eg][]{S53} is now thought to play a minor role \citep{H02},
while spirals \citep{S84, B88, S02} and interactions \citep{H06, K08}
are believed to cause more substantial changes.

The Milky Way is believed to have both a thin and a thick disc
\citep{G83, M04, J08, I08}, as do perhaps many disc galaxies
\citep{Y06}.  Relative to the thin disc, the thick disc has a higher
velocity dispersion, lags in its net rotational velocity \citep{C00},
contains older stars with lower metallicities \citep{M93} and enhanced
[$\alpha$/Fe] ratios \citep{BF05, R06, F08}.  A number of different
criteria have been used to divide stars into the two disc populations,
and some of these trends may depend somewhat on whether a spatial,
kinematic, or chemical abundance criterion is applied \citep{F08,
  S09b, L11}.  Recently, \citet{Bovy11} suggest that there is no sharp
distinction into two populations, but rather a continuous variation
in these properties.

Several models have been proposed for the formation of the thick disc:
accretion of disrupted satellite galaxies \citep{A03}, thickening of
an early thin disc by a minor merger \citep{Q93, V08}, star formation
triggered during galaxy assembly \citep{BG05, B09}, and radial
migration \citep{S09a, L10}.  While it is likely that no single
mechanism plays a unique role in its formation, our focus in this
paper is on radial migration.

As first shown by \citet{S02}, spiral activity in a disc causes stars
to diffuse radially over time.  Stars near corotation of a spiral
pattern experience large angular momentum changes of either sign that
move them to new radii without adding random motion.  As spirals are
recurring transient disturbances that have corotation radii scattered
over a wide swath of the disc, the net effect is that stars execute a
random walk in radius with a step size ranging up to $\sim 2\;$kpc,
which results in strong radial mixing.  Note that this mechanism is
distinct from the ``blurring'' caused by epicycle oscillations of
stars about their home radii, $R_h$, where $L_z^2 =
R_h^3|\partial\Phi/\partial R|_{R_h}$ in the midplane and $\Phi$
is the gravitational potential.  Radial mixing is dubbed ``churning''
and causes the home radii themselves to change.

Note that radial mixing caused by recurrent transient spirals has two
related properties that distinguish it from other processes that have
sometimes also been said to cause radial migration.  The two
distinctive properties are the absence of disc heating, already noted,
and the fact that stars mostly change places, causing little change to
the distribution of angular momentum within the disc and leaving the
surface density profile unchanged.  These are both features of
scattering at corotation; spirals also cause smaller angular momentum
changes at Lindblad resonances which do heat and spread the disc.

Bar formation causes greater angular momentum changes than those of an
individual spiral pattern \citep{F94, R98, G99, D06, M11, B11, BC11}.
However, because the bar persists after its formation, the associated
angular momentum changes have a different character from those caused
by a spiral that grows and decays quickly.  Bar formation changes the
radial distribution of angular momentum, and therefore also the
surface density profile of the disc, as well as adding a substantial
amount of radial motion \citep{Hohl71}.  Although the bar may
subsequently settle, grow, and/or slow down over time, the main
changes associated with its formation probably occur only once in a
galaxy's lifetime \citep[for a dissenting view see][]{BC02} whereas
spirals seem to recur indefinitely as long as gas and star formation
sustain the responsiveness of the disc.  Thus angular momentum changes
in the outer disc beyond the bar should ultimately be dominated by the
effects of radial mixing by dynamically-unrelated transient spirals,
as we find in this paper.  Note that \citet{M10} studied the separate
physical process of resonance overlap using steadily rotating
potentials to represent a bar and spiral, and therefore did not
address possible mixing by transient spirals in a barred galaxy model.

Both \citet{Q09} and \citet{B11} find that external bombardment of the
disc can enhance radial motion through the increase in the central
attraction and also through possible angular momentum changes.
However, the importance of bombardment is unclear since satellites may
be dissolved by tidal shocking \citep{D10} before they settle, and
furthermore satellite accretion to the inner Milky Way over the past
$\sim 10^{10}$ years is strongly constrained by the dominant old age of
thick disc stars \citep[\eg][]{W09}.

\citet{S02} discovered spiral churning in 2D simulations, and it has
been shown to occur in fully 3D simulations \citep{R08a, R08b, L10}.
\citet{S09a} present a model for Galactic chemical evolution that
includes radial mixing, and point out \citep{S09a, S09b, S11} that it
naturally gives rise to both a thin and a thick disc, under the
assumption that thick disc stars experience a similar radial churning.
\citet{L10} show that extensive spiral activity in their simulations
causes a thick disc to develop, and present a detailed comparison with
data \citep{I08} from SDSS \citep{Y00}.  \citet{SB09} and \citet{M09}
also generated galaxies with realistic break radii in the exponential
surface brightness profiles using cosmological $N$-body hydrodynamic
simulations that included radial migration.  However, \citet{B11}
report that no significant thick discs developed in their simulations.

\citet{L11} find evidence for radial mixing in the thin disc and for a
downtrend of rotational velocity with metallicity. The presence of
such a trend does not rule out migration in the thick disc, but shows
that mixing for the oldest stars cannot be complete.  \citet{Bovy11}
find a continuous change of abundance-dependent disc structure with
increasing scale height and decreasing scale length which they note is
the ``almost inevitable consequence of radial migration''.  The
decreasing metallicity gradient with age of Milky Way disc stars can
also be attributed to radial mixing \citep{Yu12}.

While \citet{L10} present circumstantial evidence for radial migration
in the thick disc, they do not show explicitly that it is occurring,
neither do they attempt to quantify the extent to which it may be
reduced by the weakened responses of thick disc stars to spiral waves
in the thin disc.  \citet{B11} find that mixing is more extensive when
spiral activity is invigorated by star formation, although the level
of spiral activity is strongly dependent on the ``gastrophysical''
prescription adopted.  They show that mixing persists even for
particles with large oscillations about the mid-plane, and they
determine migration probabilities from their simulations.

In this paper, we set ourselves the limited goal of determining the
extent of radial migration in isolated, collisionless discs with
various thicknesses and radial velocity dispersions that are subject
to transient spiral perturbations.  Following \citet{S02}, we first
present controlled simulations of two-component discs constructed so
as to support an isolated, large-amplitude spiral wave, in order to
study the detailed mechanism of migration in the separate thin and
thick discs.  We also report the responses of two-component discs to
multiple spiral patterns, both with and without a bar.

\section{Description of the simulations}
All our models use the constant velocity disc known as the Mestel disc
\citep[][hereafter BT08]{B08}.  Linear stability analysis by
\citet[][see also Zang 1976 and Evans \& Read 1998]{T81} revealed that
this disc with moderate random motion lacks any global instabilities
whatsoever when half its mass is held rigid and the centre of the
remainder is cut out with a sufficiently gentle taper.\footnote{\citet{Sell12}
finds that particle realizations of this disc are not completely stable,
but on a long time scale when $N$ is large.}   We therefore
adopt this stable model for the thin disc, and superimpose a thick
disc of active particles that has $10\%$ of the mass of the thin disc.
The remaining mass is in the form of a rigid halo, set up to ensure
that the total central attraction in the midplane is that of the
razor-thin, full-mass, untapered Mestel disc.

\subsection{Disc set up}
Ideally, we would like to select particles from a distribution
function (\DF) for a thickened Mestel disc.  \citet{T82} found a family
of flattened models that are generalizations of the razor-thin Zang
discs that have the isothermal $\hbox{sech}^2z$ vertical density
profile \citep{S42, C50}.  Unfortunately, these two-integral models
have equal velocity dispersions in the radial direction and normal to
the disc plane (see BT08), whereas we would like to set up models with
flattened velocity ellipsoids.

Since no three-integral \DF\ for a realistic disc galaxy model is
known, as far as we are aware, we start from the two-integral \DF\ for
the razor-thin disc \citep{Z76,T77}:
\begin{equation}
f_{\rm Zang}(E,L_z) \propto L_z^qe^{-E/\sigma_R^2},
\end{equation}
where $E$ is a particle's specific energy and $L_z$ its specific
$z$-angular momentum.  The free parameter $q$ is related to the
nominal radial component of the velocity dispersion through
\begin{equation}
\sigma_R = V_0(1+q)^{-1/2},
\label{qdef}
\end{equation}
with $V_0$ being the circular orbital speed at all radii.  The
value of Toomre's local stability parameter for a single component,
razor-thin Mestel disc would be
\begin{equation}
Q \equiv {\sigma_R \over \sigma_{R,{\rm min}}} = {2^{3/2}\pi \over 3.36 f(1+q)^{1/2}},
\label{Qsingle}
\end{equation}
where $f$ is the active fraction of mass in the component; note that
these expressions for both $\sigma_R$ and $Q$ are independent of
radius.  A composite model having two thickened discs, such as we
employ here, will have some effective $Q$ that is not so easily
expressed.  We choose different values of $\sigma_R$ for the thin and
thick discs, given in Table~\ref{tab:simulations} below, in order that
the thick disc has a greater radial velocity dispersion, as observed
for the Milky Way.

We limit the radial extent of the discs by inner and outer tapers
\begin{equation}
f_0(E,L_z) = \frac{f_{\rm Zang}}{[1+(L_i/L_z)^4][1+(L_z/L_o)^6]},
\end{equation}
where $L_i$ and $L_o$ are the central angular momentum values of the
inner and outer tapers respectively.  The exponents are chosen so as
not to provoke instabilities (Toomre, private communication).  We
employ the same taper function for both discs and choose $L_o =
15L_i$.  We further restrict the extent of the disc by eliminating all
particles whose orbits would take them beyond $25R_i$, where $R_i =
L_i/V_0$ is the central radius of the inner taper.  This additional
truncation is sufficiently far out that the active mass density is
already substantially reduced by the outer taper.

\begin{figure}
\includegraphics[width=84mm]{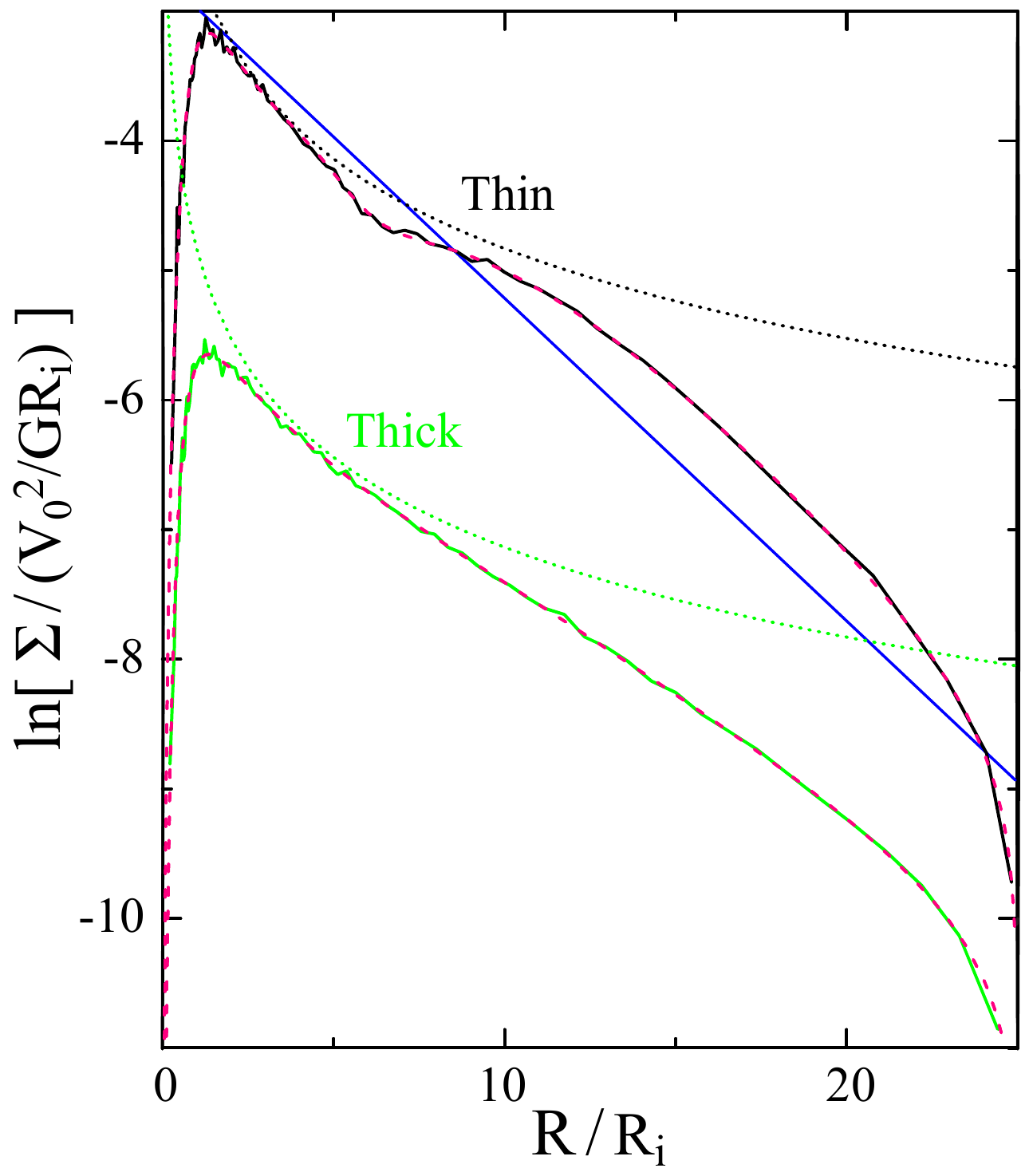}
\caption{Initial surface density profiles of the tapered thin (top black curves)
  and thick (bottom green curves) Mestel discs in simulation M2 (solid curves)
  measured from the particles.  The dashed red curves, which are
  almost perfectly overlaid by the solid curves, show the expected
  surface density of the tapered discs from integrating the \DF\ over
  all velocities, while the dotted curves indicate the density
  profiles of the corresponding non-tapered discs.  The groove in the
  thin disc is centred on $R=6.5R_i$ and is broadened by epicycle
  excursions.  The blue line of slope $-0.25$ indicates the adopted
  exponential profile of the thin disc.}
\label{fig:run94surfdn}
\end{figure}

We thicken these disc models by giving each the scaled vertical
density profile
\begin{equation}
\rho(\tilde{z}) \propto \frac{1}{\left(e^{|\tilde{z}|/2} +
  0.2e^{-5|\tilde{z}|/2}\right)^2},
\label{zprof}
\end{equation}
where $\tilde{z} = z/z_0$ with $z_0$ independent of $R$.  While both
this function and the usual $\hbox{sech}^2z$ function have harmonic
cores, we prefer eq.~(\ref{zprof}) because it approaches $\rho \propto
e^{-|\tilde z|}$ more rapidly, as suggested by data \citep[\eg][]{K02};
$z_0$ is therefore the exponential scale height when $z \gg z_0$.  We
set the vertical scale height of the thick disc to be three times that
of the thin disc, as suggested for the Milky Way \citep{J08}.

We estimate the equilibrium vertical velocity dispersion at each
$z$-height by integrating the 1D Jeans equation (BT08, eq.~4.271) in our
numerically-determined potential.  This procedure is adequate when the
radial velocity dispersion is low, but the vertical balance degrades
in populations having larger initial radial motions, which flare
outwards as the model relaxes from initial conditions, as shown below.

The tapered thin Mestel disc has a surface density that declines with
radius as shown in Fig.~\ref{fig:run94surfdn}.  While there is no
radial range that is closely exponential, we estimate an approximately
equivalent exponential radial scale length $R_d = 4.0R_i$ as shown by
the blue line.  Accordingly, we choose $z_0 = 0.4R_i$ for the thin
disc so that the ratio $R_d/z_0$ is similar to that of the Milky Way
and the average volume-corrected ratio of $7.3\pm2.2$($1\sigma$) found
by \citet{K02} for 34 nearby galaxies.  The values we adopt for each
model are given in Table~\ref{tab:simulations} below.

The radial gravitational potential gradient in the midplane of our
model is less than that of the full-mass, razor-thin, infinite Mestel
disc as a result of the reduced disc surface density, the angular
momentum tapers, the finite thickness of the discs, and gravitational
force softening.  In order to create an approximate equilibrium, we
therefore add a rigid central attraction to the self-consistent forces
from the particles in the discs at each step.  We tabulate the
supplementary central attraction needed at the initial moment to yield
a radial force per unit mass of $-V_0^2/R$ in the midplane, and apply
this unchanging extra term as a spherically symmetric, rigid central
attraction.

\Ignore{In all our models, the vertical frequency $\nu$ is about twice
  the epicyclic frequency $\kappa$ over the region where the disc
  density is high, although it decreases towards the circular angular
  frequency $\Omega = \kappa/\sqrt{2}$ elsewhere, as it should where
  the disc is negligible.  Hence, vertical resonances ($\Omega =
  \Omega_p \pm \nu/m$) in the active disc are farther from corotation
  than are the Lindblad resonances, as they should be.}

\begin{figure}
\includegraphics[width=84mm]{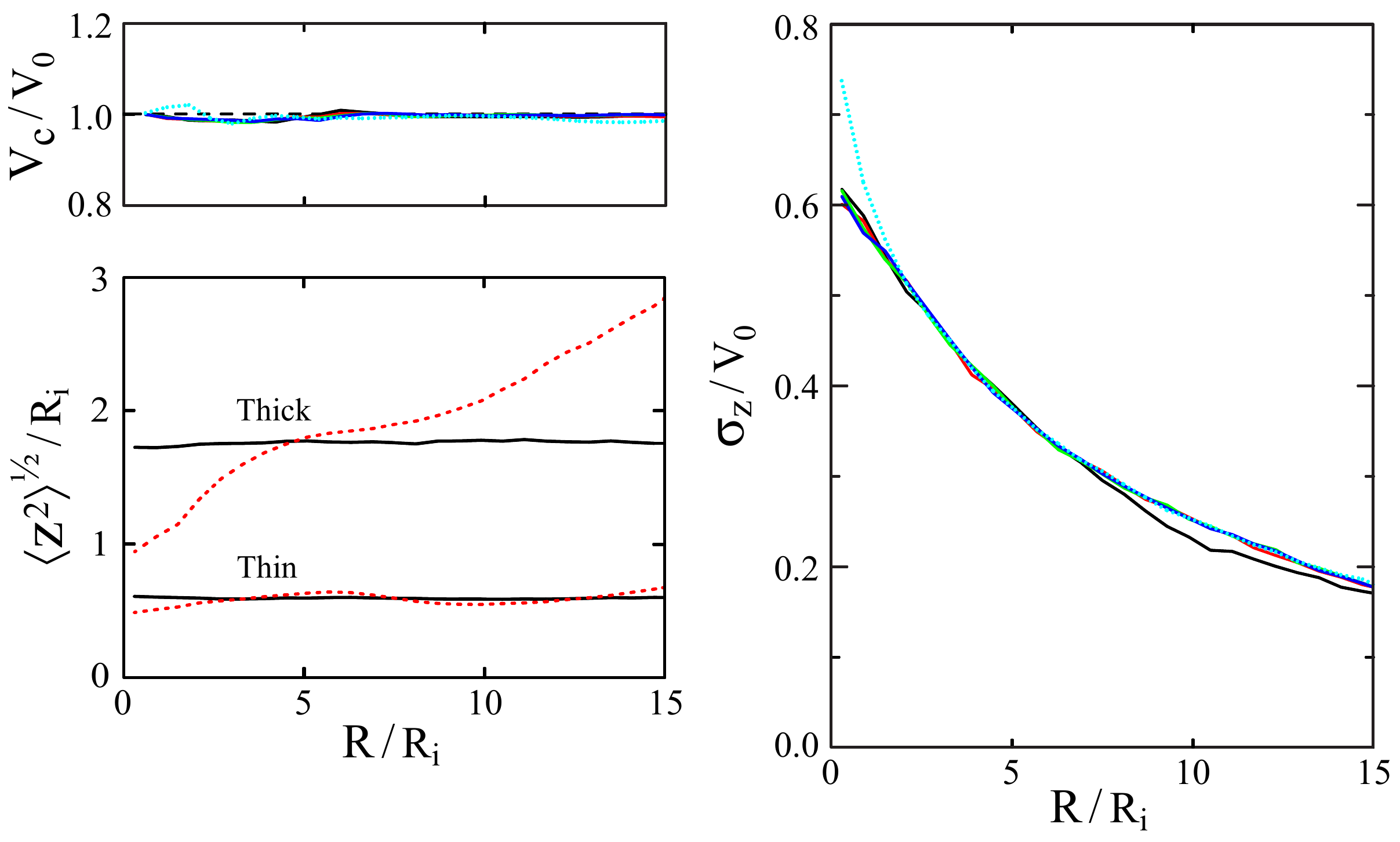}
\caption{The circular velocity in disc midplane (top left) and vertical
  velocity dispersion of the thick disc particles (right) at five
  equally spaced times during simulation M2.  The times, which include
  the initial and final moments of the simulation, are colour coded in
  temporal order by solid black, red, green, blue, and dotted cyan
  respectively.  The dashed horizontal line in the top left panel shows
  the theoretical circular velocity $V_0 = 1$.  The bottom left panel
  shows the initial $\left\langle z^2\right\rangle^{1/2}$ of the thin
  (bottom curves) and thick (top curves) discs (solid black) compared
  to that at $t = 32R_i/V_0$ (dashed red).}
\label{fig:run94vcszrmsz}
\end{figure}

\begin{table*}
\box1{
\begin{minipage}{126mm}
\caption{Parameters of all the simulations described in this paper.}
\label{tab:simulations}
\begin{tabular}{@{}lclrrrrrrc}
\hline
Simulation & Disc & $f$ & $\frac{z_0}{R_i}$ & $\frac{\sigma_R}{V_0}$ & $N$ & $m$ & $\frac{\Delta z}{R_i}$ & $\frac{\varepsilon}{R_i}$ & Rings$\times$Spokes$\times$Planes \\
\hline
M2 & thin & 0.5 & 0.4 & 0.283 & 1\,200\,000 & 0 , 2 & 0.1 & 0.15 & 120$\times$128$\times$243 \\
 & thick & 0.05 & 1.2 & 0.567 & 1\,800\,000 & & & & \\
T & thin & 0.5 & 0.4 & 0.283 & 1\,200\,000 & 0 & 0.1 & 0.15 & 120$\times$128$\times$243 \\
 & thick & 0.05 & 1.2 & 0.567 & 1\,800\,000 & & & & \\
M3 & thin & 0.3333 & 0.4 & 0.189 & 1\,200\,000 & 0 , 3 & 0.1 & 0.15 & 120$\times$128$\times$243 \\
 & thick & 0.0333 & 1.2 & 0.378 & 1\,800\,000 & & & & \\
M4 & thin & 0.25 & 0.4 & 0.142 & 1\,200\,000 & 0 , 4 & 0.1 & 0.15 & 120$\times$128$\times$243 \\
 & thick & 0.025 & 1.2 & 0.283 & 1\,800\,000 & & & & \\
M4b & thin & 0.25 & 0.2 & 0.142 & 1\,200\,000 & 0 , 4 & 0.05 & 0.075 & 120$\times$128$\times$405 \\
 & thick & 0.025 & 0.6 & 0.283 & 1\,800\,000 & & & & \\
 & massless thick & 0 & 1.2 & 0.283 & 1\,800\,000 & & & & \\
M2b & thin & 0.5 & 0.4 & 0.283 & 480\,000 & 0 , 2 & 0.1 & 0.15 & 75$\times$80$\times$625 \\
 & thick & 0.05 & 1.2 & 0.567 & 240\,000 & & & & \\
 & massless 1 & 0 & 0.5 & 0.567 & 240\,000 & & & & \\
 & massless 2 & 0 & 0.6 & 0.567 & 240\,000 & & & & \\
 & massless 3 & 0 & 0.7 & 0.567 & 240\,000 & & & & \\
 & massless 4 & 0 & 0.8 & 0.567 & 240\,000 & & & & \\
 & massless 5 & 0 & 1.6 & 0.567 & 240\,000 & & & & \\
 & massless 6 & 0 & 2.0 & 0.567 & 240\,000 & & & & \\
 & massless 7 & 0 & 2.4 & 0.567 & 240\,000 & & & & \\
M2c & thin & 0.5 & 0.4 & 0.283 & 480\,000 & 0 , 2 & 0.1 & 0.15 & 75$\times$80$\times$243 \\
 & thick & 0.05 & 1.2 & 0.567 & 240\,000 & & & & \\
 & massless 1 & 0 & 1.2 & 0.378 & 240\,000 & & & & \\
 & massless 2 & 0 & 1.2 & 0.454 & 240\,000 & & & & \\
 & massless 3 & 0 & 1.2 & 0.680 & 240\,000 & & & & \\
 & massless 4 & 0 & 1.2 & 0.794 & 240\,000 & & & & \\
 & massless 5 & 0 & 1.2 & 0.907 & 240\,000 & & & & \\
 & massless 6 & 0 & 1.2 & 1.021 & 240\,000 & & & & \\
 & massless 7 & 0 & 1.2 & 1.134 & 240\,000 & & & & \\
TK & thick & 0.5 & 1.2 & 0.283 & 1\,800\,000 & 0 , 2 & 0.1 & 0.15 & 120$\times$128$\times$243 \\
UC , UCB1, & thin & 0.4 & 0.4 & 0.227 & 200\,000 & $\le$ 8, $\ne$ 1 & 0.2 & 0.3 & 75$\times$80$\times$125 \\
UCB2 & thick & 0.04 & 1.2 & 0.454 & 300\,000 & & & & \\
\hline
\end{tabular}

\medskip
The first column gives the simulation designation used in the text.
The next five columns give the properties of each disc, one for each
line: $f$ is the mass fraction, $z_0$ its vertical scale height,
$\sigma_R$ is the nominal radial velocity dispersion, and $N$ is the
number of particles in the disc.  The final four columns give $m$,
the active sectoral harmonic(s), $\Delta z$ the vertical spacing of
the grid planes, $\varepsilon$ the softening length, and the grid size.
The ``massless'' discs of simulations M4b, M2b, and M2c describe
massless thick discs composed of test particles.  $z_0$, $\Delta z$,
and $\varepsilon$ are in terms of $R_i$, and $\sigma_R$ is in terms of
the circular velocity $V_0$.
\end{minipage}
}
\kern3.8cm\box1
\end{table*}

We find that our model is close to equilibrium, with an initial virial
ratio of the particles of $\approx 0.52$.  This value adjusts quickly
to reach a steady virial ratio of $0.50$ within the first 32 dynamical
times (defined below).  As noted above, the initial imbalance seems to
arise mostly from the vertical velocity structure, for which we adopted
the 1D Jeans equation.  The adjustment of the model to equilibrium is
illustrated in Fig.~\ref{fig:run94vcszrmsz}; the bottom left panel
shows that thicknesses of both discs change as the system relaxes,
increasing in the outer parts and decreasing in the inner parts --
notice that the change in thickness is least over the radius range
$2R_i \la R \la 10R_i$, where the surface mass density is closest to its
untapered value (dotted curves in Fig.~\ref{fig:run94surfdn}).  The
changes, which are larger for the radially-hotter thick disc, result
from the radial excursions of the particles, which may take them far
from their initial radii for which the vertical velocity was set.
However, neither the radial balance (top left panel) nor the vertical
velocity dispersion (right panel) of the thick disc change
significantly from their initial values.  After this initial relaxation
from initial conditions, we do not observe any significant flaring.
A negligibly small fraction ($<1\%$) of particles escape from our grid
by the end of the simulation, and most particle loss takes place as
the model settles.

Following \cite{S02}, we seed a vigorously unstable spiral mode by
adding a Lorentzian groove in angular momentum to the \DF\ of the thin
disc only:
\begin{equation}
f(E,L_z) = f_0(E,L_z)\left[1 + \frac{\beta w_L^2}{(L_z - L_*)^2 + w_L^2}\right].
\label{eqgroove}
\end{equation}
Here $\beta$, a negative quantity, is the depth of the groove, $w_L$
is its width, and $L_*$ is the angular momentum of the groove centre.
This change to the \DF\ seeds a predictable spiral instability
\citep{S91}.  The groove in the thin disc has the parameters $\beta =
-0.9$, $w_L = 0.3R_iV_0$, and $L_* = 6.5R_iV_0$.  We find that a deeper
and wider groove is needed than that used by \citet{S02} in order to
excite a strong spiral in our thickened disc.

Since a groove of this kind will provoke instabilities at many
sectoral harmonics, $m$, we restrict disturbance forces from the
particles to a single non-axisymmetric sectoral harmonic to prevent
other modes from growing.  Corotation for a global spiral mode excited
by this groove is at a radius somewhat greater than $L_*/V_0$, where
local theory would predict.  A large-scale mode is not symmetric about
the groove centre due to the geometric variation of surface area with
radius -- an effect that is neglected in local theory.  However, the
radius of corotation approaches the local theory prediction for modes
of smaller spatial scale, \ie\ as $m \rightarrow \infty$.

As in \cite{S02}, we choose the circular velocity $V_0$ to be our unit
of velocity, $R_i$ our unit of length, the time unit or dynamical time
is $\tau_0 = R_i/V_0$, our mass unit is $M_0 = V_0^2R_i/G$, and $L_i$
is our unit of angular momentum.  One possible scaling to physical
units is to choose $R_i = 0.75\;$kpc, with $R_d = 4.0R_i$ being the
equivalent scale length of the thin disc, and $\tau_0 = 3.0\;$Myr,
leading to $V_0 = 244\;$km~s$^{-1}$ and $M_0 = 1.04 \times
10^{10}\;$M$_\odot$.

\subsection{Numerical procedure}
We use the 3D polar grid described in \cite{S97} to determine the
gravitational field of the particles.  This ``old-fashioned'' method
is not only well suited to the problem, but also has the advantage of
being tens of times faster than the ``modern'' methods recently
reviewed by Dehnen \& Read (2011).  We employ a grid having $120$
rings, $128$ spokes, and $243$ vertical planes, and adopt the cubic
spline softening rule recommended by \citet{M92}, which yields the
full attraction of a point mass at distances greater than two
softening lengths ($2\varepsilon$).  The Plummer rule used in
\cite{S02} is suited for razor-thin discs where it mimics the effect
of disc thickness but, because it weakens inter-particle forces on all
scales, it is unsuited for 3D simulations where disc thickness is
already included in the particle distribution.  The value of
$\varepsilon$, given in Table~\ref{tab:simulations}, exceeds the
vertical grid spacing in order to minimize the grid dependence of the
inter-particle forces.

In the simulations described in \S\S3\&4, we use quiet starts
\citep{S86} for both discs to reduce the initial amplitude of the
seeded unstable spiral mode far below that expected from shot noise.
In 3D, this requires many image particles for each fundamental
particle: we place two at each $(R,\phi,z)$ position with oppositely
directed $z$-velocities and two more reflected about the midplane at
the point $(R,\phi,-z)$.  We then place images of these four particles
at equal intervals in $\phi$, each set having identical velocity
components in cylindrical polar coordinates.  For these simulations,
the thin disc has $1\,200\,000$ particles and the thick disc
$1\,800\,000$.  All the particles in a single population have equal
mass, but particle masses differ between populations in order to
create the desired ratio of disc surface densities.

We evaluate forces from the particles at intervals of $0.02\tau_0$,
and step forward some of the particles at each evaluation.  Since the
orbital periods of particles span a wide range, we integrate their
motion when $R > 2R_i$ using longer time steps in a series of five
zones with the step doubling in length for each factor 2 in radius
\citep{S85}.  We also subdivide the time step for particles within
$R = 0.5R_i$ without updating forces, with further decreases by a factor
of $2$ for every factor of $2$ decrease in radius \citep{S04}.  Note
that very few particles have $R < 0.5R_i$, which is well within the
inner taper where the rigid component of the force dominates.  Tests
with shorter time steps yielded similar results.

\subsection{List of simulations}
For convenience, we summarize the parameters of all the simulations
presented in this paper in Table~\ref{tab:simulations}.  Simulation T
is constrained to remain axisymmetric, while all those whose
identifier begins with `M' are highly controlled experiments designed
to support a single spiral instability.  We first present, in \S3, a
detailed description of M2, which supports a bisymmetric spiral, and
briefly compare it to simulation T.  Variants of M2, with many
populations of test particles are presented in \S3.4, while simulation
TK, which has a thick disc only, is described in \S3.5.  Simulations
that support a single spiral of higher angular periodicity are
motivated and described in \S4.  The last three simulations in the
Table, with identifier beginning with `U', are uncontrolled
experiments presented in \S5 that explore the consequences of multiple
spirals, with and without a bar.

\section{A single bisymmetric spiral}
Our first objective is to study radial migration in both the thick and
thin discs due to a single spiral disturbance.  We therefore present
simulation M2, which is designed to support an isolated $m=2$ spiral
instability.

We measure
\begin{equation}
A_m(t) = 2\pi \int_{R_1}^{R_2} \Sigma(R,\phi,t) e^{im\phi} \; dR,
\label{danlampl}
\end{equation}
where $\Sigma(R,\phi,t)$ is the vertically-integrated mass surface
density of the particles at time $t$, and we generally choose
$R_1 = 1.5R_i$ and $R_2 = 19R_i$.

\begin{figure}
\includegraphics[width=84mm]{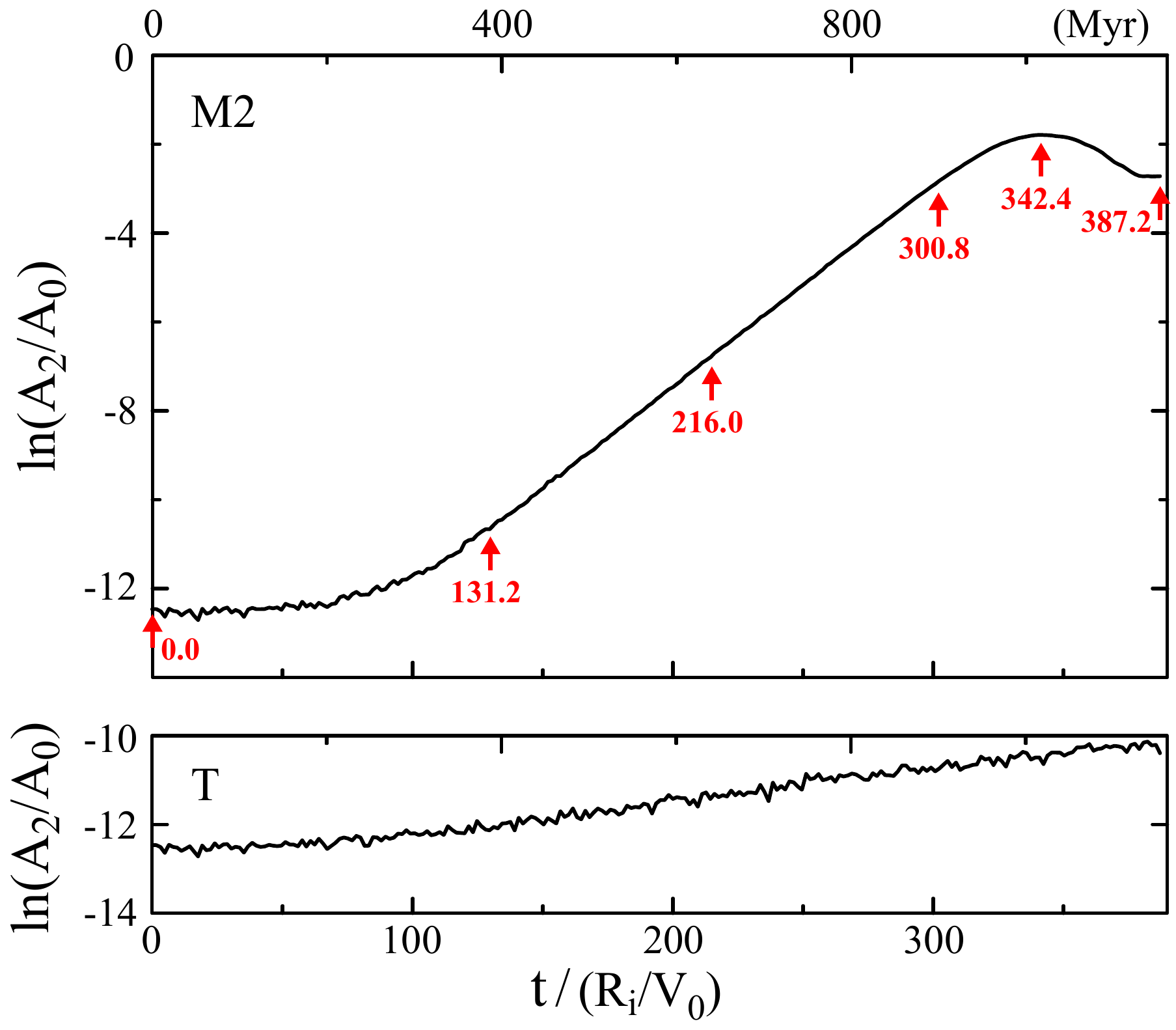}
\caption{The time evolution of $A_2/A_0$ in simulations M2 (top) and T
  (bottom).  Note the difference in the vertical scales.  The numbered
  arrows in the top plot mark the six times of M2 at which the
  snapshots of the discs are shown in Fig.~\ref{fig:run94pnts}.  The
  top axis shows time scaled to physical units using the adopted scaling
  at the end of section 2.1.}
\label{fig:run9496ampl}
\end{figure}

The top panel of Fig.~\ref{fig:run9496ampl} shows the time evolution
of $A_2/A_0$, revealing that the mode grows exponentially, peaks, and
then decays non-exponentially to a trough.  The peak is $2.55$ times
higher than the minimum of the following trough.  Continued evolution
reveals that the amplitude rises again due to the growth of a
secondary wave.  In order to isolate the single initial spiral, we
stop the simulation at the trough and compare quantities, such as the
specific angular momenta of particles, at this final time with those
at the initial time $t = 0.0R_i/V_0$.  The bottom panel plots the same
quantity from simulation T in which disturbance forces from the
particles were restricted to the axisymmetric ($m=0$) term; the very
slow rise is caused by the gradual degradation of the quiet start.

\begin{figure}
\includegraphics[width=84mm]{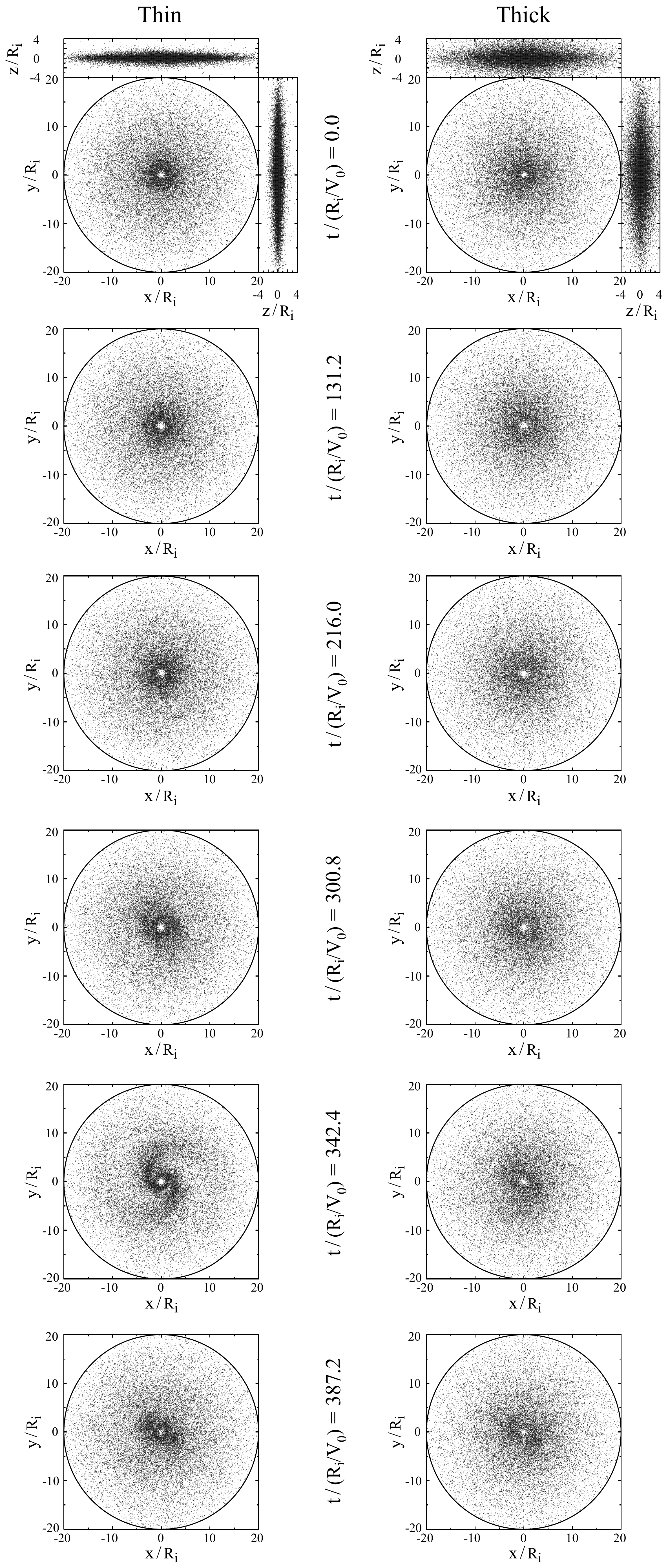}
\caption{Snapshots of the thin (left column) and thick (right column)
  discs in simulation M2 at the six times marked in
  Fig.~\ref{fig:run9496ampl}.  One in every $24$ particles is plotted
  for the thin disc, and one in every $36$ for the thick.  The side
  views are shown only at the initial time because they change little
  throughout the simulation.}
\label{fig:run94pnts}
\end{figure}

Fig.~\ref{fig:run94pnts} shows snapshots of the thin and thick discs
of M2 at six different times of the simulation that are marked by the
red arrows in Fig.~\ref{fig:run9496ampl}.

We find the period of exponential growth can be very well fitted by a
single mode, using the apparatus described in \cite{S86}.
Table~\ref{tab:spirals} reports the fitted pattern speed $\Omega_p$
and growth rate $\gamma$, together with other parameters of the spiral
in M2.  Furthermore, the non-linear evolution visible in
Fig.~\ref{fig:run94pnts} shows no evidence of a bar.  Thus, the
selected time period of the simulation M2 presents the opportunity to
study radial migration due to a single, well isolated spiral wave.
Note that the spiral pattern makes a full rotation every
$2\pi/\Omega_p \simeq 45.5R_i/V_0$.

\begin{table*}
\begin{minipage}{126mm}
\caption{Measured values from fits to the spiral modes.}
\label{tab:spirals}
\begin{tabular}{@{}lccccccc}
\hline
 & $m$ & $\Omega_p/\frac{V_0}{R_i}$ & $\gamma$ & $\frac{R_c}{R_i}$ & $\left(\frac{A_m}{A_0}\right)_{\rm peak}$ & P/T & $t_p/\frac{R_i}{V_0}$ , ($Myr$) \\
\hline
M2 & 2 & 0.138 & 0.0454 & 7.24 & 0.168 & 2.55 & 73.5 , 221 \\
M3 & 3 & 0.147 & 0.0417 & 6.81 & 0.099 & 2.15 & 76.9 , 231 \\
M4 & 4 & 0.150 & 0.0361 & 6.67 & 0.063 & 1.80 & 76.3 , 229 \\
M4b & 4 & 0.151 & 0.0465 & 6.63 & 0.083 & 3.07 & 81.3 , 244 \\
M2b & 2 & 0.138 & 0.0457 & 7.25 & 0.166 & 2.52 & 72.1 , 216 \\
M2c & 2 & 0.138 & 0.0456 & 7.25 & 0.166 & 2.52 & 74.4 , 223 \\
TK & 2 & 0.135 & 0.0180 & 7.40 & 0.064 & 1.68 & 149 , 446 \\
\hline
\end{tabular}

\medskip
The second column gives the angular periodicity of the spiral or the
greatest active sectoral harmonic $m$ in the simulation.  The next
three columns give the pattern speed $\Omega_p$, growth rate $\gamma$,
and the corotation radius $R_c$ of the spiral obtained by fitting it's
exponential growth using the method in \cite{S86}.  The next two
columns provide the spiral's peak amplitude and
$\left(\frac{A_m}{A_0}\right)_{\rm
  peak}/\left(\frac{A_m}{A_0}\right)_{\rm trough}$.  And the last column
gives the duration of the peak amplitude, which we measure between the
moment the amplitude reaches the trough and the moment preceding the
peak at which the amplitude is at the same level as the trough amplitude.
The second number in the peak duration column is the time in physical
units using the adopted scaling at the end of section 2.1.
\end{minipage}
\end{table*}

\begin{figure}
\includegraphics[width=71mm]{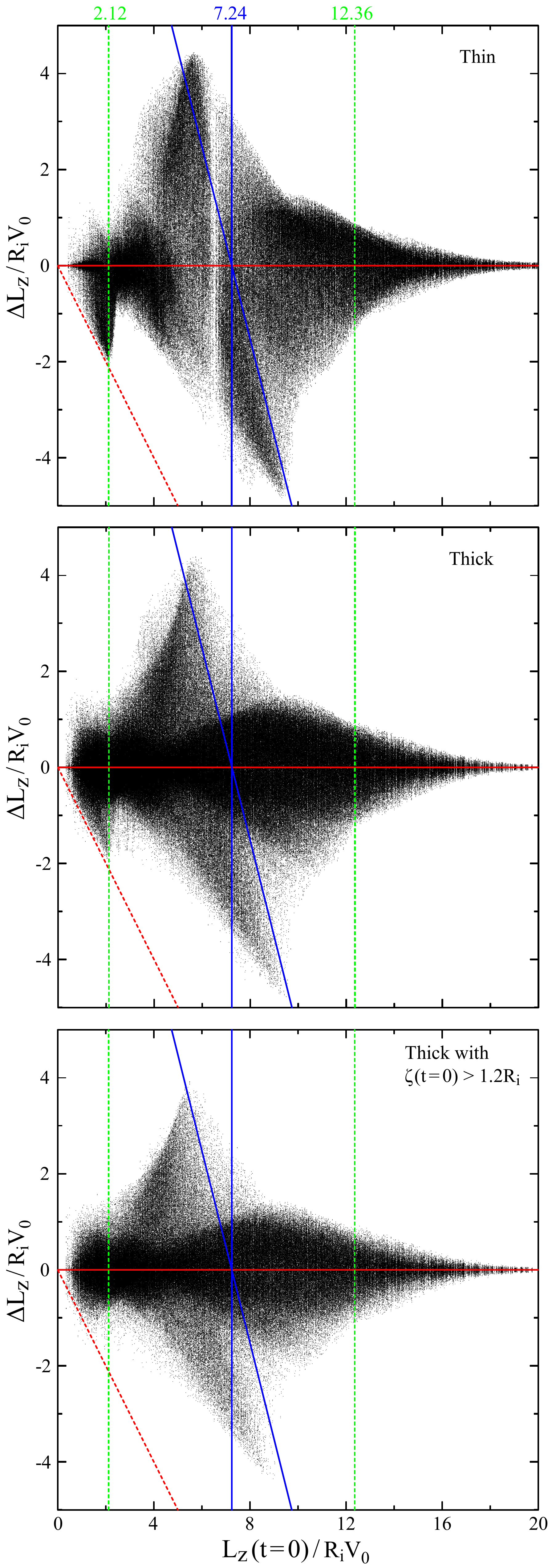}
\caption{Angular momentum changes of particles in M2 as a function of
  their initial angular momenta.  The top panel is for thin disc
  particles, the middle for all thick disc particles, and the bottom
  for thick disc particles having notional vertical amplitude $\zeta >
  1.2R_i$.  The horizontal red line denotes zero change.  The vertical
  lines mark the Lindblad resonances (dashed green) and corotation
  (solid blue).  The solid blue line with a slope of $-2$ illustrates
  the locus of particles whose changes would be symmetric about
  corotation.  The dashed red line of slope $-1$ shows the
  $\Delta L_z = -L_z(t=0)$ locus below which particles end up on
  retrograde orbits.}
\label{fig:run94lch}
\end{figure}

\subsection{Angular momentum changes}
Fig.~\ref{fig:run94lch} shows the change in the specific $z$-angular
momenta $\Delta L_z$ of the particles in the thin disc (top) and
the thick disc (middle) of simulation M2 against their initial
$L_z$.  The deficiency of particles at $L_z(t=0) = L_* = 6.5R_iV_0$ in the
top panel is due to the groove in the thin disc.  The vertical
lines show the locations of the corotation (solid blue) and the Lindblad
resonances (dashed green) for nearly circular orbits.

The maximum changes in angular momentum occur near corotation, and lie
close to the solid blues line of slope $-2$; these particles cross
corotation, in both directions, to about the same radial distance away
from it as they were initially.  As for the razor-thin disc, we find
that large angular momentum changes occur only around the time that
the spiral saturates, for reasons explained by \citet{S02}.  While
many particles near the inner Lindblad resonance lose angular
momentum, only a tiny fraction end up on retrograde orbits.  Note also
that $\rmsLz \approx 9.0\times 10^{-6}R_iV_0$ in simulation T, where
non-axisymmetric forces were eliminated, showing that changes in $L_z$
due to orbit integration and noise errors are tiny in comparison to
those caused by the spiral.

\begin{figure}
\includegraphics[width=84mm]{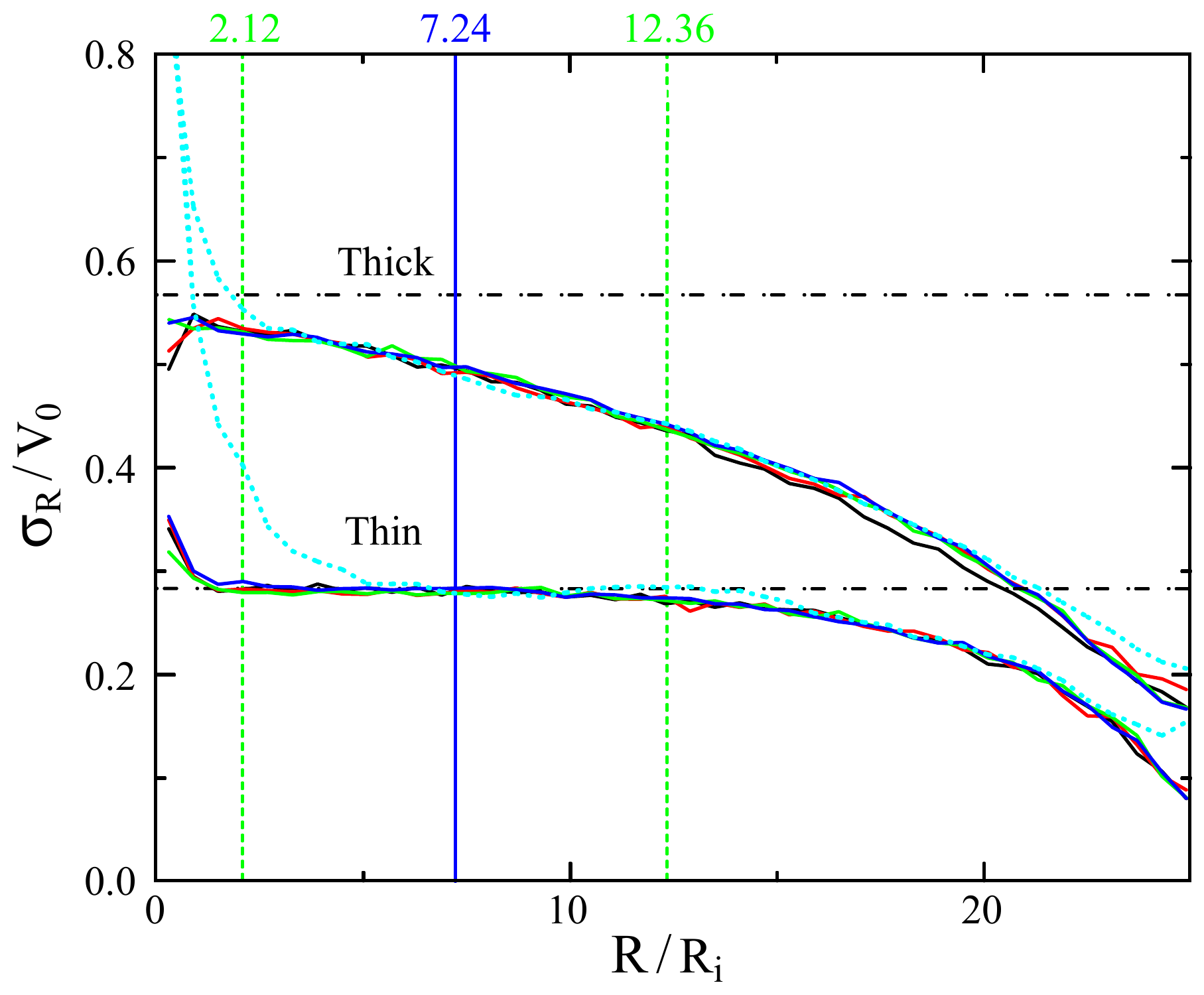}
\caption{The radial variations of $\sigma_R$ in the thin (bottom curves)
  and thick (top curves) discs of M2 at the same five different times as
  used in the top left and right panels of Fig.~\ref{fig:run94vcszrmsz} with
  the same line colour and style coding.  The horizontal dot-dashed curves
  show the theoretical initial dispersions from the untapered discs, while
  the vertical lines mark the principal resonances of the spiral, colour and
  style coded as in Fig.~\ref{fig:run94lch}.}
\label{fig:run94sigmar}
\end{figure}

For each particle, we determine the initial value of
\begin{equation}
Z = \frac{1}{2}v_z^2 + \frac{1}{2}\nu^2z^2,
\label{Ez}
\end{equation}
where $v_z$ is the vertical velocity component at distance $z$ from
the midplane and $\nu$ is the vertical frequency measured in the
midplane at the particle's initial radius $R$.  For particles whose
vertical and radial oscillations are small enough to satisfy the
epicycle approximation, $Z=E_{z,\rm epi}$ the energy of its vertical
oscillation; the vertical potential is roughly harmonic for $|z| \la
0.4R_i$.  Even though the epicycle approximation is not satisfied for
the majority of particles, we compute an initial notional vertical
amplitude
\begin{equation}
\zeta = \left( \frac{2Z}{\nu^2} \right)^{1/2},
\label{zmax}
\end{equation}
for them all.  The value of $\zeta$ defined in this way, i.e.\ at the
initial moment only, yields a convenient approximate ranking of the
vertical oscillation amplitudes of the particles, although it is clear
that in most cases $\zeta < z_{\rm max}$, the maximum height a
particle may reach.  Note that since each disc component in our models
has an initial thickness that is independent of radius, the
distribution of $\zeta$ is independent of $L_z(t=0)$.

Notice from Fig.~\ref{fig:run94lch} that changes for the thick disc
particles are only slightly smaller than those for thin disc
particles.  In order to emphasize this point, the bottom panel
displays only those thick disc particles for which $\zeta > 1.2R_i$,
or one scale height of the thick disc, revealing that even for these
particles changes can be almost as large as those in the thin disc.

Fig.~\ref{fig:run94sigmar} shows the evolution of the radial velocity
dispersion for the thin and thick discs.  As expected \citep{S02}, the
large angular momentum changes near corotation cause little heating.
Some heating occurs near the Lindblad resonances, and is greater near
the inner resonance, again as expected.

\begin{figure}
\includegraphics[width=84mm]{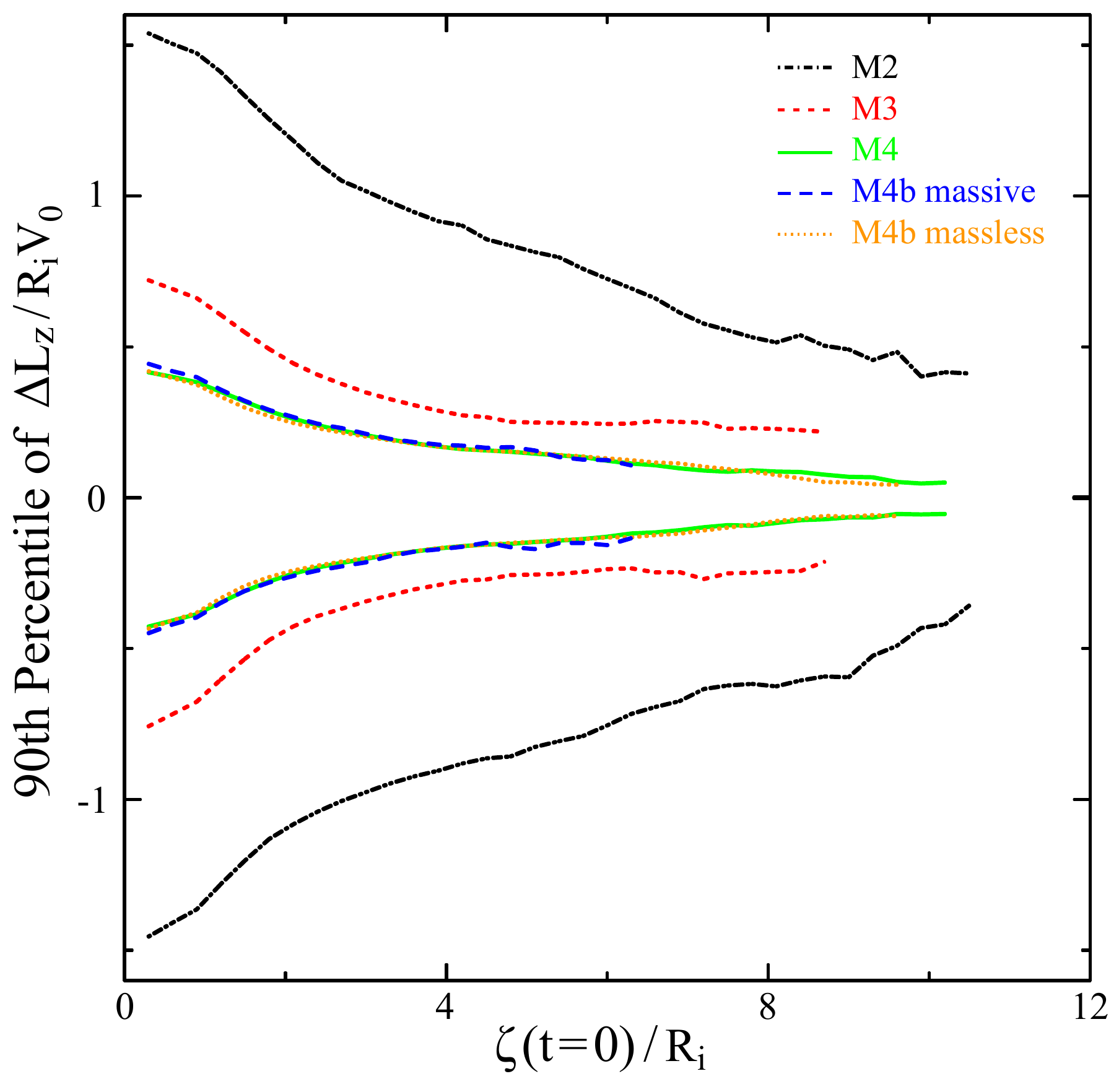}
\caption{The 5th and 95th percentiles of the angular momentum changes
  in the thick discs of simulations M2 (dot-dashed black), M3
  (short-dashed red), M4 (solid green), the massive thick disc of M4b
  (long-dashed blue), and the massless thick disc of M4b (dotted orange)
  as a function of notional vertical amplitude $\zeta$.  Although we have
  smoothed the curves, the rising noise with increasing $\zeta$ is caused
  by the decrease in the number of particles per bin.  The curves stop when
  the number of particles per bin drops below twenty.}
\label{fig:run949799101lchmx}
\end{figure}

\subsection{Distribution of angular momentum change}
Table~\ref{tab:rmslch} lists the root mean square, maximum positive,
and maximum negative changes in angular momentum for both all the
particles and only those that satisfy $\zeta > z_0$ in each disc.

\begin{figure}
\includegraphics[width=84mm]{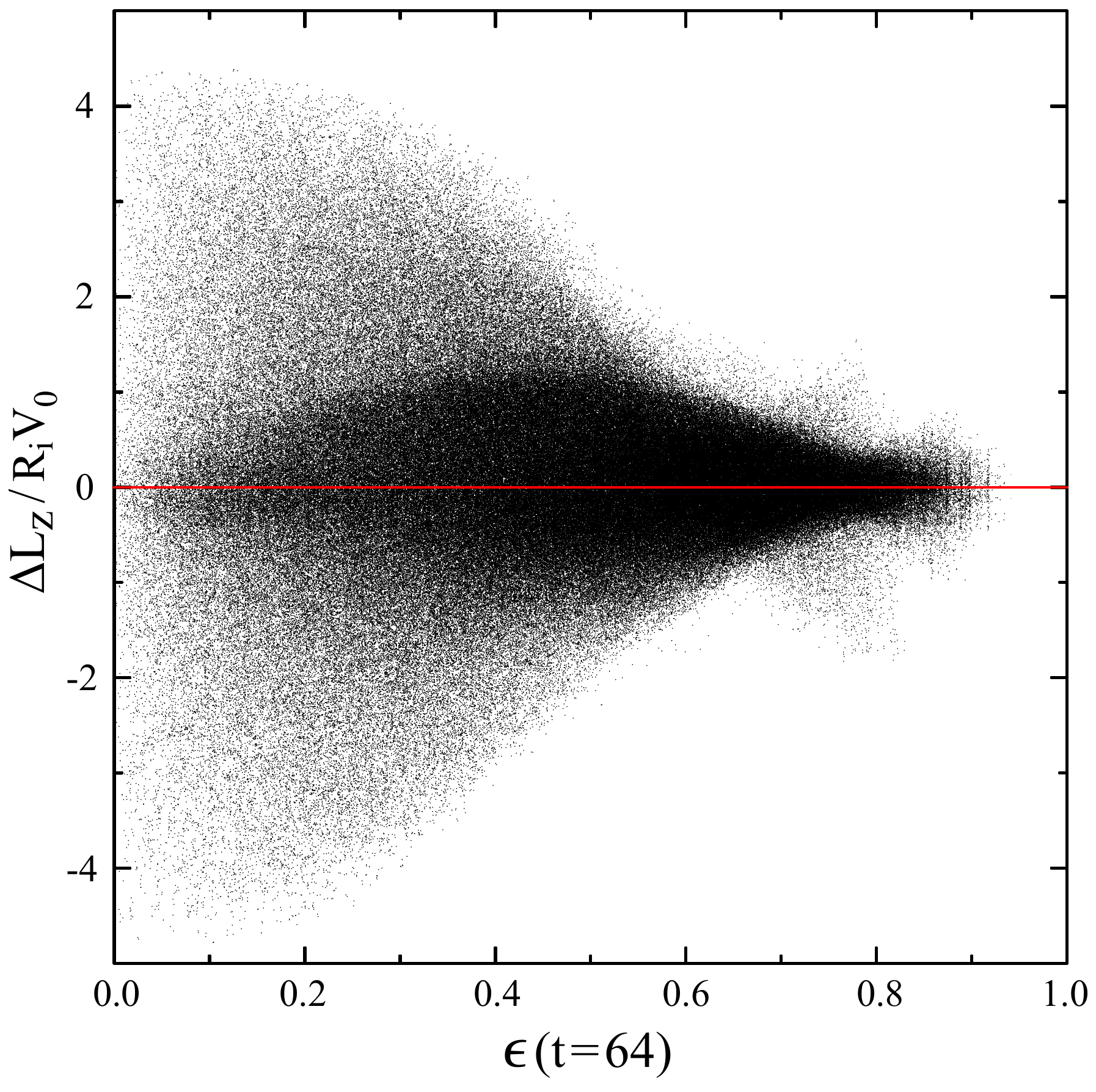}
\caption{Angular momentum changes of thick disc particles between the
  final time and $t = 64R_i/V_0$ having
  $2.5R_iV_0 \le L_z(t=64) \le 10.0R_iV_0$ in simulation M2 as a function
  of their eccentricities at $t = 64R_i/V_0$.  The plot for the thin disc
  is quite similar.  The red horizontal line shows zero angular momentum
  change.}
\label{fig:run94lchecc}
\end{figure}

\begin{table}
\caption{Changes in angular momentum resulting from the spirals and bar.}
\label{tab:rmslch}
\begin{tabular}{@{}lcccc}
\hline
 & Disc & $\frac{\rmsLz}{R_iV_0}$ & $\frac{\textrm{Max }\Delta L_z}{R_iV_0}$ & $\frac{\textrm{Max -}\Delta L_z}{R_iV_0}$ \\
\hline
M2 & thin & 1.59 , 1.56 & 4.44 , 4.37 & -4.83 , -4.79 \\
   & thick & 0.97 , 0.85 & 4.38 , 3.91 & -4.78 , -4.46 \\
M3 & thin & 0.92 , 0.89 & 2.79 , 2.72 & -2.67 , -2.61 \\
   & thick & 0.54 , 0.43 & 2.76 , 2.22 & -2.60 , -2.12 \\
M4 & thin & 0.62 , 0.60 & 2.03 , 1.97 & -2.11 , -2.02 \\
   & thick & 0.37 , 0.30 & 1.93 , 1.65 & -1.99 , -1.59 \\
M4b & thin & 0.80 , 0.78 & 2.25 , 2.18 & -2.17 , -2.08 \\
    & thick & 0.47 , 0.40 & 2.21 , 1.84 & -2.09 , -1.83 \\
    & massless & 0.38 , 0.27 & 2.17 , 1.43 & -2.05 , -1.48 \\
M2b & thin & 1.58 , 1.54 & 4.45 , 4.36 & -4.57 , -4.43 \\
    & thick & 0.95 , 0.84 & 4.33 , 3.76 & -4.52 , -4.02 \\
    & massless 1 & 1.08 , 1.06 & 4.42 , 4.32 & -4.43 , -4.43 \\
    & massless 2 & 1.06 , 1.02 & 4.41 , 4.19 & -4.54 , -4.22 \\
    & massless 3 & 1.04 , 0.98 & 4.41 , 4.08 & -4.47 , -4.27 \\
    & massless 4 & 1.02 , 0.95 & 4.48 , 4.11 & -4.43 , -4.15 \\
    & massless 5 & 0.89 , 0.72 & 4.35 , 3.46 & -4.40 , -3.67 \\
    & massless 6 & 0.83 , 0.64 & 4.39 , 3.11 & -4.29 , -3.64 \\
    & massless 7 & 0.79 , 0.58 & 4.05 , 3.06 & -4.30 , -3.37 \\
M2c & thin & 1.57 , 1.53 & 4.46 , 4.35 & -4.58 , -4.44 \\
    & thick & 0.96 , 0.85 & 4.33 , 3.76 & -4.50 , -4.03 \\
    & massless 1 & 1.19 , 1.06 & 4.37 , 3.75 & -4.42 , -4.11 \\
    & massless 2 & 1.08 , 0.95 & 4.37 , 3.81 & -4.46 , -4.01 \\
    & massless 3 & 0.86 , 0.76 & 4.37 , 3.66 & -4.49 , -4.09 \\
    & massless 4 & 0.80 , 0.72 & 4.35 , 3.70 & -4.28 , -4.09 \\
    & massless 5 & 0.75 , 0.68 & 4.22 , 3.79 & -4.33 , -3.89 \\
    & massless 6 & 0.72 , 0.65 & 4.18 , 3.56 & -4.51 , -3.81 \\
    & massless 7 & 0.69 , 0.63 & 4.21 , 3.51 & -4.29 , -3.81 \\
TK & thick & 0.78 , 0.75 & 3.37 , 3.26 & -2.88 , -2.77 \\
UC & thin & 2.65 , 2.31 & 13.97 , 11.60 & -10.43 , -9.48 \\
   & thick & 1.95 , 1.75 & 12.71 , 10.33 & -10.15 , -10.04 \\
UCB1 & thin &  3.15 , 2.76 & 16.42 , 12.41 & -13.22 , -11.12 \\
     & thick & 2.34 , 2.10 & 16.00 , 13.80 & -12.00 , -10.92 \\
UCB2 & thin & 3.53 , 3.46 & 19.23 , 19.14 & -16.74 , -16.74 \\
     & thick & 2.54 , 2.28 & 19.31 , 17.40 & -15.83 , -13.63 \\
\hline
\end{tabular}

\medskip
For each disc of every simulation, the first numbers in the third, fourth,
and fifth columns give the root mean square, maximum positive, and maximum
negative changes in angular momentum respectively.  We calculate these for
all the particles except those that escaped the grid for which the initial
angular momentum lies in an interval of the spiral's main influence around
its corotation.  This interval, in values in terms of $R_iV_0$, is
$[2.5,10]$ for simulations M2, M2b, and M2c, $[4.5,9.5]$ for M3, $[5.0,8.5]$
for M4 and M4b, and $[3.5,9.5]$ for TK.  We do not confine $L_z(t=0)$ to
such an interval for simulations UC, UCB1, and UCB2 since the influence of
their spirals and bar span almost the entire range.  The second numbers in
the last three columns give the same results but only for particles having
notional vertical amplitude $\zeta > z_0$.
\end{table}

Fig.~\ref{fig:run949799101lchmx} shows the 5th and 95th percentile
values of the changes in angular momentum as a function of initial
notional vertical amplitude, $\zeta$ (eq.~\ref{zmax}), using bin
widths of $0.3R_i$.  The affect of radial migration seems to decrease
almost linearly with increasing $\zeta$.

Radial migration should also be weaker for particles having larger
radial oscillations or epicycles of large amplitude.  As reasoned by
\citet{S02}, particles on more eccentric orbits cannot hold station
with a steadily rotating spiral, because their angular velocities vary
significantly as they oscillate radially.  Fig.~\ref{fig:run94lchecc},
which is for only those particles in the angular momentum range
$2.5R_iV_0 \le L_z(t=64) \le 10.0R_iV_0$, confirms that the largest
angular momentum changes occur among particles having the least
eccentric orbits.  \footnote{The energy cut-off we apply (see after
  eq.~4) eliminates the most eccentric orbits from this plot.}  It
shows angular momentum changes from, and eccentricities at, time $t =
64R_i/V_0$, which is after the model has settled from its mild initial
imbalance, but before any substantial angular momentum changes have
occurred.  We define eccentricity as $\epsilon = (R_{\rm a}-R_{\rm
  p})/(R_{\rm a}+R_{\rm p})$, in which $R_{\rm a}$ and $R_{\rm p}$ are
respectively the initial apo-centre and peri-centre distance of the
orbit of a particle having the same $L_z$, but whose motion is
confined to the midplane of the axisymmetric potential.

\subsection{Effect of radial migration on vertical oscillations}
In the previous subsection, we showed how the particles' angular
momentum changes vary with initial amplitude of vertical motion.
Here, we present the converse: how the vertical oscillations are
affected by the radial excursions.

Since the notional amplitude of vertical motion $\zeta$
(eq.~\ref{zmax}) is accurate only in the epicycle approximation, we
determine a particle's actual maximum vertical excursion, $z_{\rm
  max}$, by integrating its motion in a frozen, azimuthally-averaged
potential for many radial periods.  We do this twice for each particle
in simulation M2, at time $t = 64R_i/V_0$ starting from the particle's
phase space coordinates in the frozen potential at that moment and
again at the final time $t = 387.2R_i/V_0$.

\begin{figure}
\includegraphics[width=80mm]{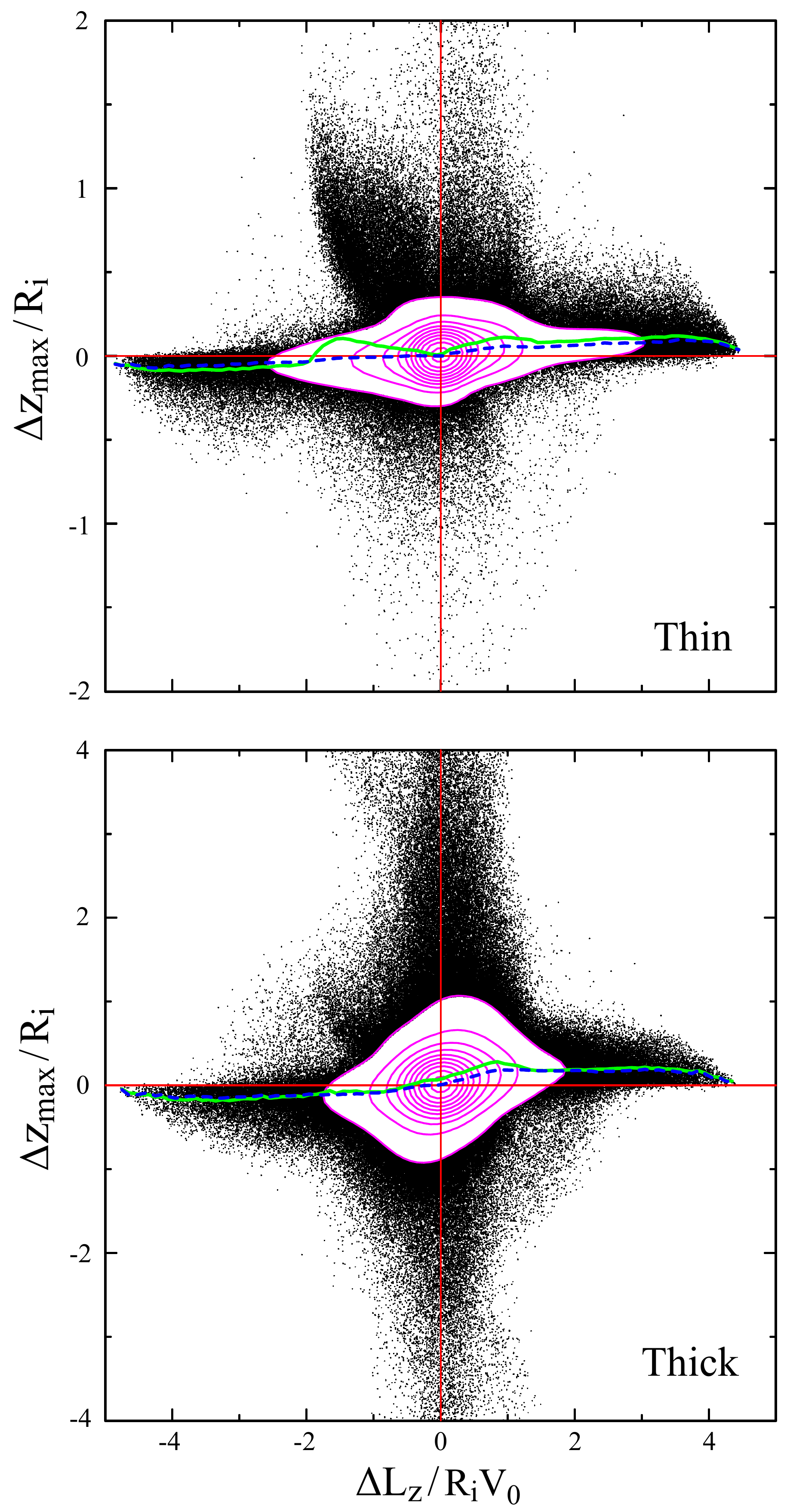}
\caption{Changes in maximum vertical excursions of particles in the
  thin (top) and thick (bottom) discs of M2 versus changes in their
  angular momenta.  Contours are linearly spaced in number density and
  we plot only those points that lie outside the lowest contour.  The
  mean (solid green) and median (dashed blue) show systematic
  variations, as expected.  Note that the vertical scales differ in
  the two plots.}
\label{fig:run94zmaxchlch}
\end{figure}

Fig.~\ref{fig:run94zmaxchlch} shows that, on average, the vertical
excursions of particles increase for those that move radially outwards
and decrease for those that move inwards.  This is expected, because
restoring forces to the mid-plane are weaker at larger radii.  For
both discs, the mean (solid green) and median (dashed blue) curves
show a roughly constant $\Delta z_{\rm max}$ for a wide range of
$\Delta L_z$.  For the thick disc, this constant $\Delta z_{\rm max}$
is about twice as great as that for the thin.

While the large majority of the particles lie in the contoured region,
the outliers exhibit significant substructure.  The dense group of
points near a line of slope of $-1$ in the second quadrant are
particles with initial home radii within and near the inner $m=2$
vertical resonance\footnote{Vertical resonances occur when $m(\Omega -
  \Omega_p) = \pm\nu$}, which lies at the radius $2.23R_i$, not far
from the radial inner Lindblad resonance at $2.12R_i$.  Usually $\nu
\gg \kappa$, which causes vertical resonances to be found
significantly farther from corotation than the radial resonances, but
in our case the inner taper reduces the inner surface density so that
$\nu \sim \kappa$ in this part of the disc.  Thus, these particles are
scattered vertically at the inner vertical resonance at the same time
as they lose angular momentum at the radial inner Lindblad resonance.
Another outlying group can be seen in the first quadrant, with large
$\Delta z_{\rm max}$ for small $\Delta L_z$; these particles have
quite eccentric initial orbits that have particular radial phases just
before the spiral saturates.  Either they are at their pericentres and
lie near corotation just trailing either spiral arm, or they are at
their apocentres and surround the outer ends of the spiral arms.
Being at these special locations when the spiral wave is strongest,
gives them instantaneous angular frequencies about the centre that
cause them to experience more nearly steady, and not oscillatory,
torques from the perturbation, leading to some angular momentum gain.
Thus these particles move onto even more eccentric orbits and their
increased $z_{\rm max}$ occurs at their new larger apocentres.

\begin{figure}
\includegraphics[width=84mm]{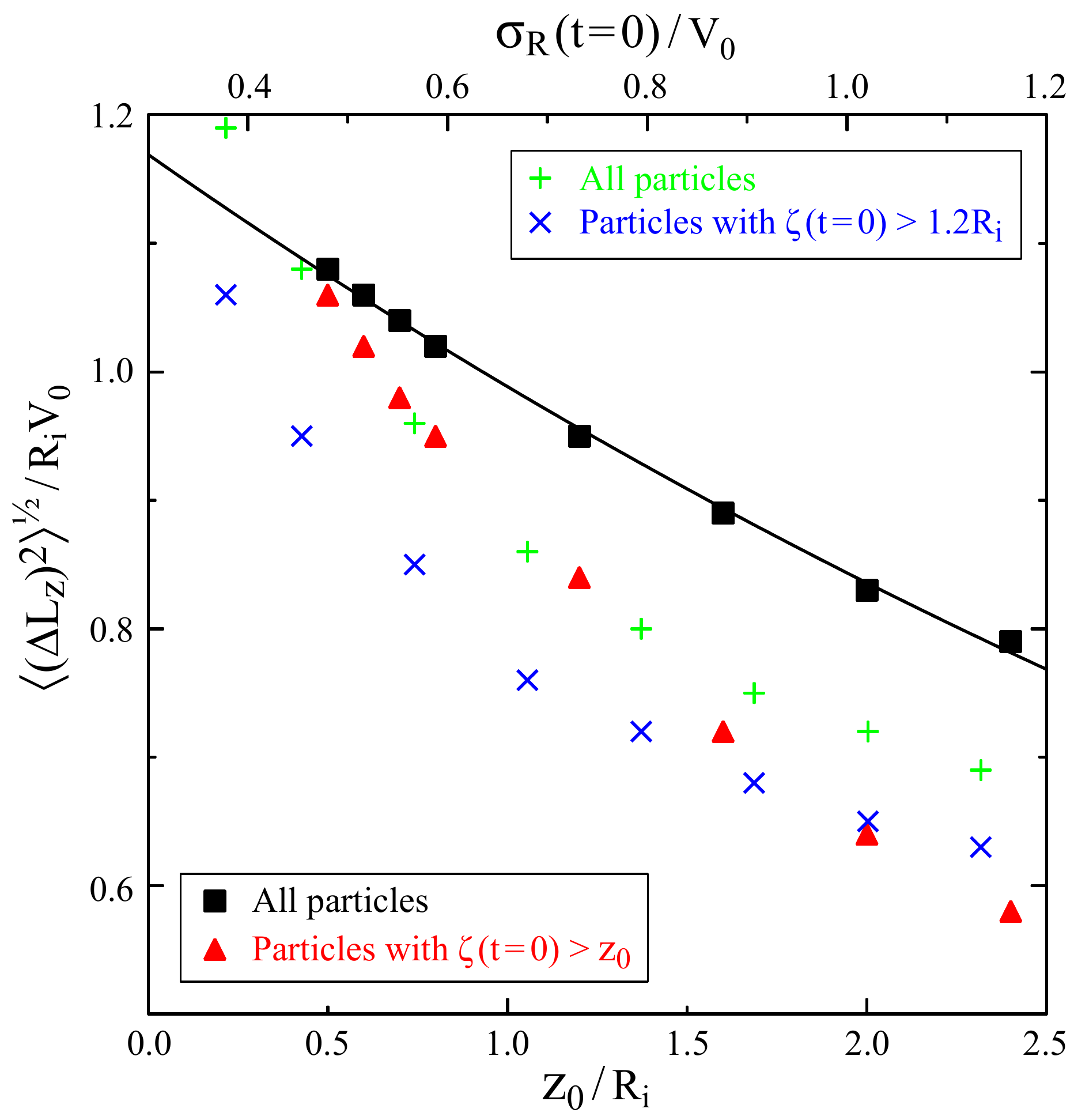}
\caption{The filled symbols show $\rmsLz$ as a function of vertical
  thickness (bottom axis) of particle populations of simulation M2b.
  The black squares are for all the particles in the angular momentum
  range $2.5R_iV_0 \le L_z(t=0) \le 10.0R_iV_0$, while the red
  triangles are for only those particles having $\zeta > z_0$.  The
  line is a least-squares fit to the black squares of the form $\rmsLz
  \propto e^{-z_0/5.95R_i}$.  The green plus symbols show the
  variation of $\rmsLz$ with initial radial velocity dispersion (top
  axis) of all the particles in the same $L_z(t=0)$ range from
  populations in simulation M2c.  The blue crosses are for only those
  with $\zeta > z_0 = 1.2R_i$.}
\label{fig:run102103rmslch}
\end{figure}

\subsection{Effects of disc thickness and radial velocity dispersion}
Our somewhat surprising finding from simulation M2 is that angular
momentum changes in the thick disc are only slightly smaller than
those in the thin, and also in the razor-thin disc of \citep{S02}.
Despite the fact that thick disc particles both rise to greater $z$
heights and have larger epicycles, on average, than do thin disc
particles, we observe only a mild decrease in their response to spiral
forcing.  This finding suggests that the potential variations of a
spiral having a large spatial scale, such as the $m=2$ spiral mode in
simulation M2, couple well to particles having large vertical motions
and epicycle sizes.

To provide more detailed information about how the extent of angular
momentum changes vary with disc thickness, we added seven test
particle populations to some simulations.  In M2b, all test particle
populations have the same initial radial velocity dispersion as the
massive thick discs, but have scale heights in the range
$0.5R_i \leq z_0 \leq 2.4R_i$.  Simulation, M2c, employs seven test
particle populations having the same scale height ($z_0 = 1.2R_i$)
as the massive thick disc, but with differing $\sigma_R$.

Being test particles, they do not affect the dynamics of the spiral
instability, but merely respond to the potential variations that arise
from the instability in the massive components.  These simulations have
identically the same physical properties as M2 but lower numerical
resolution as summarized in Table~\ref{tab:simulations}; the reduced
numerical resolution remains adequate since the fitted spiral mode is
little changed from that in simulation M2 (Table~\ref{tab:spirals}).

Variations of $\rmsLz$ with both disc thickness, at fixed radial
velocity dispersion, and of radial dispersion at fixed thickness, are
displayed in Fig.~\ref{fig:run102103rmslch}.  The decrease is somewhat
more rapid in subpopulations of particles that start with a notional
vertical amplitude $\zeta > z_0$, as seems reasonable.  The fitted
line indicates that $\rmsLz$ decays approximately exponentially with
disc thickness with a scale that can be related to theory -- see
\S\ref{theocomp}.  Table~\ref{tab:rmslch} gives both the plotted root
mean square, as well as the maximum positive, and maximum negative
$\Delta L_z$ for all the particles of each population and separately
for those with $\zeta > z_0$ of each disc.

We caution that the information in Table~\ref{tab:rmslch} and in
Fig.~\ref{fig:run102103rmslch} quantifies how the {\em responsiveness}
of a test particle population scales when subject to a fixed
perturbation.  Self-consistent spiral perturbations may differ in
strength, spatial scale, and/or time dependence causing quite
different angular momentum changes.

\subsection{Thick disc only}
For completeness, we also present simulation TK, which has a single,
half-mass active disc with a substantial thickness.  We inserted the
same initial groove, and restricted forces to $m=2$ only.

It is interesting that a groove in the thick disc still creates a
spiral instability, but one that grows less rapidly and saturates at a
lower amplitude than we found in M2.  As a consequence, the spread
in $\rmsLz$ is about $1/2$ as large as in M2.  Again we find that the
radial migration is reduced by increasing disc thickness, but not
inhibited entirely.

\section{Other simulations}
The potential of a plane wave disturbance in a thin sheet having a
sinusoidal variation of surface density $\Sigma_a$ in the
$x$-direction is
\begin{equation}
\label{eq:prtbn}
\Phi_a(x,z) = -{2\pi G \Sigma_a \over |k|} e^{ikx} e^{-|kz|}
\end{equation}
where $k$ is the wavenumber (BT08, eq.~5.161).  The exponential decay
away from the disc plane is steeper for waves of smaller spatial
scale, \ie\ larger $|k|$.  While spirals are not simple plane waves in
a razor-thin sheet, this formula suggests that we should expect radial
migration to be weakened more by disc thickness for spirals of smaller
spatial scale or higher angular periodicity -- note that the azimuthal
wavenumber, $k_\phi = m/R$.  The simulations in this section study the
effect of changing this parameter.

\Ignore{
Thus the extent of radial migration should depend upon at least three
factors: the spatial scale of the spiral relative to the disc
thickness, the peak amplitude of the spiral, $\Sigma_a$, and probably
also its time dependence.}

\subsection{Spirals of different angular periodicities}
We wish to create spiral disturbances that are similar to that in M2,
but have higher angular periodicities.  Since we expect migration to
depend not only on the spatial scale relative to the disc thickness,
but also the peak amplitude, $\Sigma_a$, and perhaps also the time
dependence, we try to keep as many factors unchanged as possible.

The mechanics of the instability seeded by a groove depends strongly
on the supporting response of the surrounding disc \citep{S91}, which
in turn, according to local theory, depends on the vigour of the swing
amplifier \citep{T81}.  Thus to generate a similar spiral disturbance
with $m>2$, we need to hold the key parameters $X$ and $Q$ at similar
values.  For an $m$-armed disturbance in a thin, single component
Mestel disc with active mass fraction $f$, the locally-defined
parameter
\begin{equation}
X = \frac{2}{fm}
\end{equation}
is independent of radius.  We therefore scale the active mass
fractions in both components as $f \propto m^{-1}$ -- recall that we
used a half-mass disc for $m=2$.  In order to preserve the same $Q$
value (eq.~\ref{Qsingle}), the radial velocity dispersion of the
particles also has to be reduced as $\sigma_R \propto f$.

Simulations M3 and M4 therefore have lower surface densities and
smaller velocity dispersions in order to support similarly growing
spiral modes of sectoral harmonic $m=3$ and 4 respectively.  A further
simulation M4b is described below.  These models are all seeded with a
groove of the same form (eq.~\ref{eqgroove}) and parameters as that in
M2.

Note that the instability typically extends between the Lindblad
resonances, which move closer to corotation as $m$ is increased.  In
the Mestel disc, the radial extent of the mode varies as
\begin{equation}
\frac{R_{\rm OLR}}{R_{\rm ILR}} = \frac{m + \sqrt{2}}{m-\sqrt{2}},
\end{equation}
\ie\ $R_{\rm OLR}/R_{\rm ILR} \approx 5.8$, 2.8 \& 2.1, for $m
=2$, 3 \& 4 respectively.  Since we have also decreased the in-plane
random motion in proportion to the surface density decrease, the
decreased size of the in-plane epicycles somewhat compensates for
the smaller scale of the mode, although the ratio is not exactly
preserved.

\subsection{Results for $m>2$}
Table~\ref{tab:spirals} gives our estimates of the spiral properties
in each simulation; uncertainties in the measured frequencies are
typically $\la 2\%$ \citep{S86}.  The pattern speeds of these
instabilities do not change much with $m$, except that we find the
radius of corotation lies closer to the groove centre $L_*/V_0 =
6.5R_i$, as expected.  The table also gives the time during which the
amplitude of the wave is equal to or greater than that at the
post-peak trough, which again does not vary much with angular
periodicity.  However, both the growth rates and the peak amplitudes
of the modes decrease from M2, to M3 and M4.

Table~\ref{tab:rmslch} includes the root mean square, maximum
positive, and maximum negative angular momentum changes for M3, M4,
and M4b, measured in each case to the moment at which $A_m$ passes
through the first minimum after the mode has saturated.  Comparison
with M2 reveals that increasing $m$ causes a roughly proportionate
decrease in the angular momentum changes, in part because the
saturation amplitude is lower, but perhaps also because the spatial
scale is reduced.  Note that again some thick disc particles in both
M3 and M4 have $\Delta L_z$ values almost as large as the greatest in
the thin disc, as was also the case for M2.

Fig.~\ref{fig:run949799101lchmx} shows 5th and 95th percentile values
of $\Delta L_z$ versus initial notional vertical amplitude $\zeta$ for
simulations M3, M4, and M4b.  Compared to the curve of M2, the $\Delta
L_z$ values are smaller for greater $m$ -- \ie\ the extent of radial
mixing is substantially lessened.  Note that this difference could
have a variety of causes, such as the lower limiting amplitude of the
mode, or possibly the different growth rate of the mode, and/or the
different disc thickness relative to the spatial scale of the mode.

This last factor is one we are able to eliminate.  The thickness of
the discs of simulations M2, M3, and M4 were held fixed as we
increased $m$ and reduced the surface density.  In order to eliminate
a change in the ratio of disc thickness to spatial scale of the mode,
we ran a further simulation M4b with the same in-plane parameters as
in M4, but with half the disc thickness.  We also halved the gravity
softening length and the vertical spacing of the grid planes.

This change restores the growth-rate of the mode in run M4b to a value
quite comparable to that in simulation M2 (Table~\ref{tab:spirals}).
The saturation amplitude, while larger than in simulation M4, is still
about half that in M2.  In addition, we added a test particle
population to simulation M4b that has parameters identical to those of
the massive thick disc, including $\sigma_R$ or $q$, except its
vertical scale height $z_0$ is that of the thick disc of M2, M3, and
M4.  Again as expected, Table~\ref{tab:rmslch} reveals that $\rmsLz$
is substantially lower in the massless disc than in the thinner,
massive disc.

\subsection{Comparison with theory}
\label{theocomp}
In order to make sense of these results, we here compare with the
theoretical picture developed by \citet{S02}.

First, we eliminate the possibility that the spiral in the simulations
with higher $m$ is ``on'' for too long for optimal migration.
\citet{S02} argue that efficient mixing by the spiral requires the
duration of the peak amplitude be less than half the period of a
horseshoe orbit, so that each particle experiences only a single
scattering.  They show that the minimum period of a horseshoe orbit
varies as $|\Psi_0|^{-1/2}$, where the potential amplitude of the
spiral perturbation at corotation varies with the spiral density
amplitude, $\Sigma_a$, and sectoral harmonic as $|\Psi_0| \propto
\Sigma_a/m$ (eq.~\ref{eq:prtbn}).  Thus the weaker density amplitude
that we find with higher $m$ implies that the minimum periods of the
horseshoe orbits are greater, and the condition for efficient mixing
is more strongly fulfilled for $m>2$.

Theory also suggests that an $m$-dependence of the peak amplitude is
unavoidable if the spiral saturates due to the onset of many horseshoe
orbits, as was argued in \citet{S02}.  In their notation, orbits
librate -- \ie\ are horseshoes -- when $E_p < p^2$, where the
frequency $p \propto m|\Psi_0|^{1/2}$.  Since $|\Psi_0| \propto
\Sigma_a/m$ (eq.~\ref{eq:prtbn}), we see that $p^2 \propto m\Sigma_a$,
suggesting that as $m$ increases, horeshoe orbits become important at
a lower peak density.  Comparing M2 with M4b, we find
(Table~\ref{tab:spirals}) the relative limiting amplitudes $\Sigma_a
\propto m^{-1}$, which is consistent with the idea that the spiral
instability saturates when the importance of horseshoe orbits reaches
very nearly the same level.  The fixed thickness and softening length
used in M2, M3 and M4 disproportionately weakens the potential of the
spiral as $m$ rises, and spoils this exact scaling.

Note also that both $\rmsLz$ and the extreme values measured from M4b
are almost exactly half those in M2, which is also consistent with the
horseshoe orbit theory developed by \citet{S02}.  Since they showed
that the maximum $\Delta L_z \propto |\Psi_0|^{1/2}$, the relations in
the previous paragraph require $\Delta L_{z,\rm max} \propto m^{-1}$
as we observe.  As the rms value also scales in the same way, it would
seem the entire distribution of $\Delta L_z$ scales with $m$ in the
same way.  Again, the constant disc thickness prevents this prediction
from working perfectly for M3 and M4.

We stress that this scaling holds because we took some care to ensure
the key dynamical properties of the disc were adjusted appropriately.
Spirals in a disc having a different responsiveness would have both a
different growth rate and probably also peak amplitude, and the
behaviour would not have manifested a simple $m$-dependence.

Finally, we motivate the exponential fit to the variation of $\rmsLz$
with $z_0$, for the same $m=2$ spiral disturbance.  We have already
shown that $\rmsLz \propto |\Psi_0|^{1/2}$, and have argued that the
spiral potential decays away from the mid-plane as $e^{-|kz|}$
(eq.~\ref{eq:prtbn}), with $k_\phi = m/R$.  Na\"\i vely, we could set
$z = z_0$, $k=k_\phi = m/R$, $m=2$, and $R=R_c$, since angular
momentum changes are centred on corotation, leading to $\rmsLz \propto
e^{-z_0/R_c}$.  The fitted scale is $5.95R_i$, which is somewhat
smaller than $R_c = 7.24R_i$.  The dominant cause of this discrepancy is
probably that spiral is not a plane wave, curvature is important for
$m=2$, and that its wavenumber is larger than $k_\phi$ because the
spiral ridges are inclined to the radial direction; we should
therefore expect the potential to decay away from the mid-plane rather
more rapidly, in the sense that we measure.

We conclude that angular momentum changes scale with spiral amplitude
in the manner predicted in \citet{S02} and, furthermore, the limiting
amplitude itself is determined by their theory.  The variation with
disc thickness is also in the sense expected from the theory, but the
quantitative prediction is not exact.

\section{Unconstrained Simulations}
Having studied at length the effects of a single spiral wave, we now
wish to illustrate the effects of multiple spirals.  We present three
simulations, UC, UCB1, and UCB2, that have no initial groove and the
initial positions of the particles are random (\ie\ not a quiet start)
since we wish spirals to develop quickly from random fluctuations.  We
include gravitational disturbance forces from all sectoral harmonics
$0 \leq m \leq 8$, except $m=1$, which we omit in order to avoid
imbalanced forces from a possibly asymmetric distribution of particles
in a rigid halo with a fixed centre.

All the many simulations of this type that we have run have ultimately
developed a strong bar at the centre.  We here compare radial
migration in three separate cases: simulation UC avoids a bar for a
long period and supports many transient spirals, while simulations
UCB1 and UCB2 have identical numerical parameters but form a bar quite
early.  In all three cases, the combined thin and thick discs have a
smaller active mass fraction, $f=0.44$, compared with $f=0.55$ for
simulation M2.  A smaller active mass and a larger gravitational
softening length help to delay bar formation but also weaken the $m=2$
spiral amplitudes somewhat.  The numerical parameters of these
simulations are also listed in Table~\ref{tab:simulations}.  Since bar
formation in these models is stochastic \citep[\eg][]{SD09}, the
different bar-formation times simply arise from different initial
random seeds.

\subsection{Multiple spirals only}
\begin{figure}
\includegraphics[width=84mm]{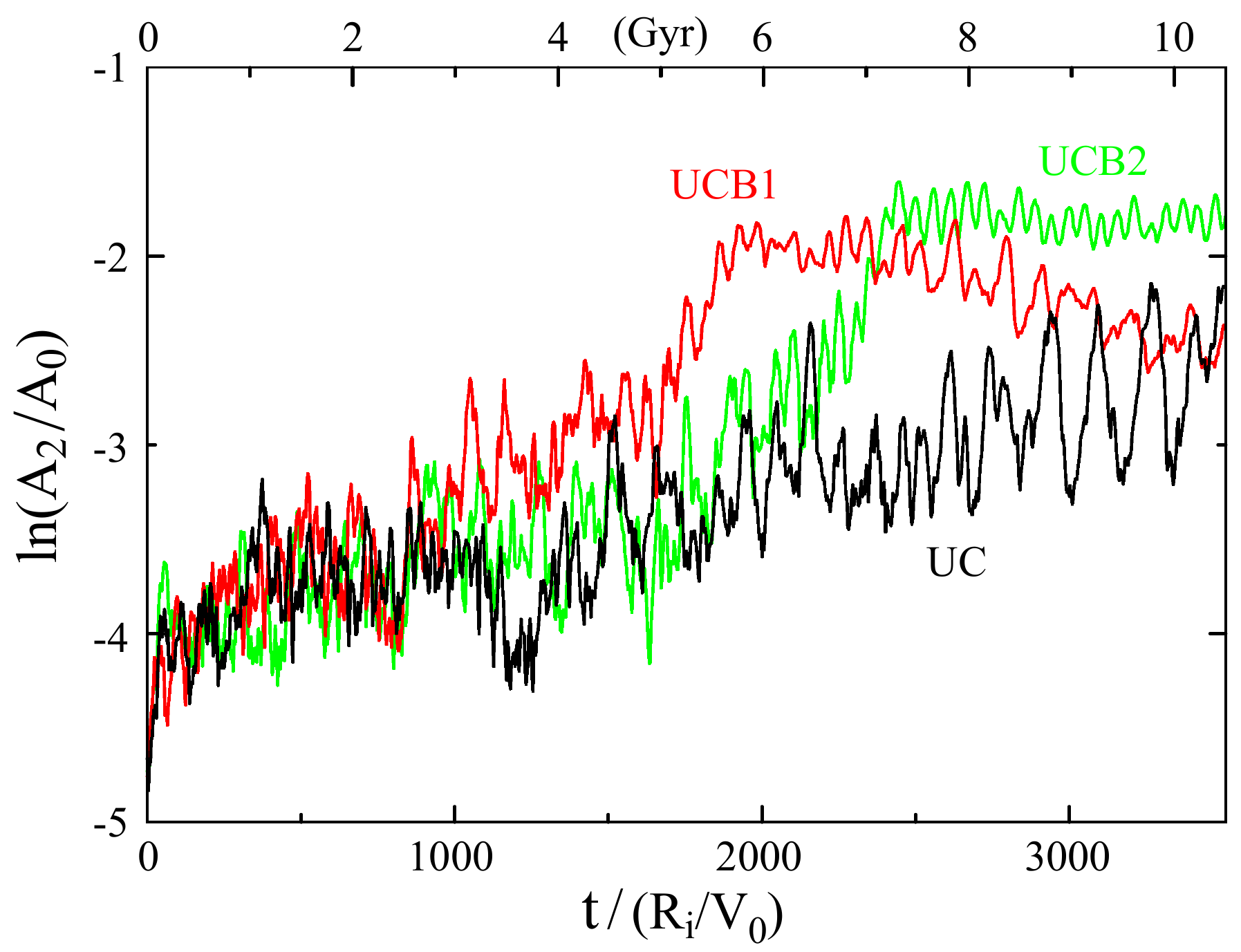}
\caption{The time evolution of $A_2/A_0$ in simulations UC (black),
  UCB1 (red), and UCB2 (green).  The top axis shows time scaled to
  physical units using the adopted scaling.}
\label{fig:run104ampl}
\end{figure}

The black curve in Fig.~\ref{fig:run104ampl} shows the evolution of
$A_2/A_0$ in simulation UC.  The spiral amplitudes rise to the point
at which significant particle scattering begins by $t \sim
1\,500R_i/V_0$ and we present the behaviour up to time $3\,500R_i/V_0$
shortly before a bar begins to form.  The period $1\,500R_i/V_0 < t <
3\,500R_i/V_0$ during which particles are scattered by spiral activity
corresponds to $\sim 6.0\;$Gyr with our adopted scaling.

\begin{figure}
\includegraphics[width=84mm]{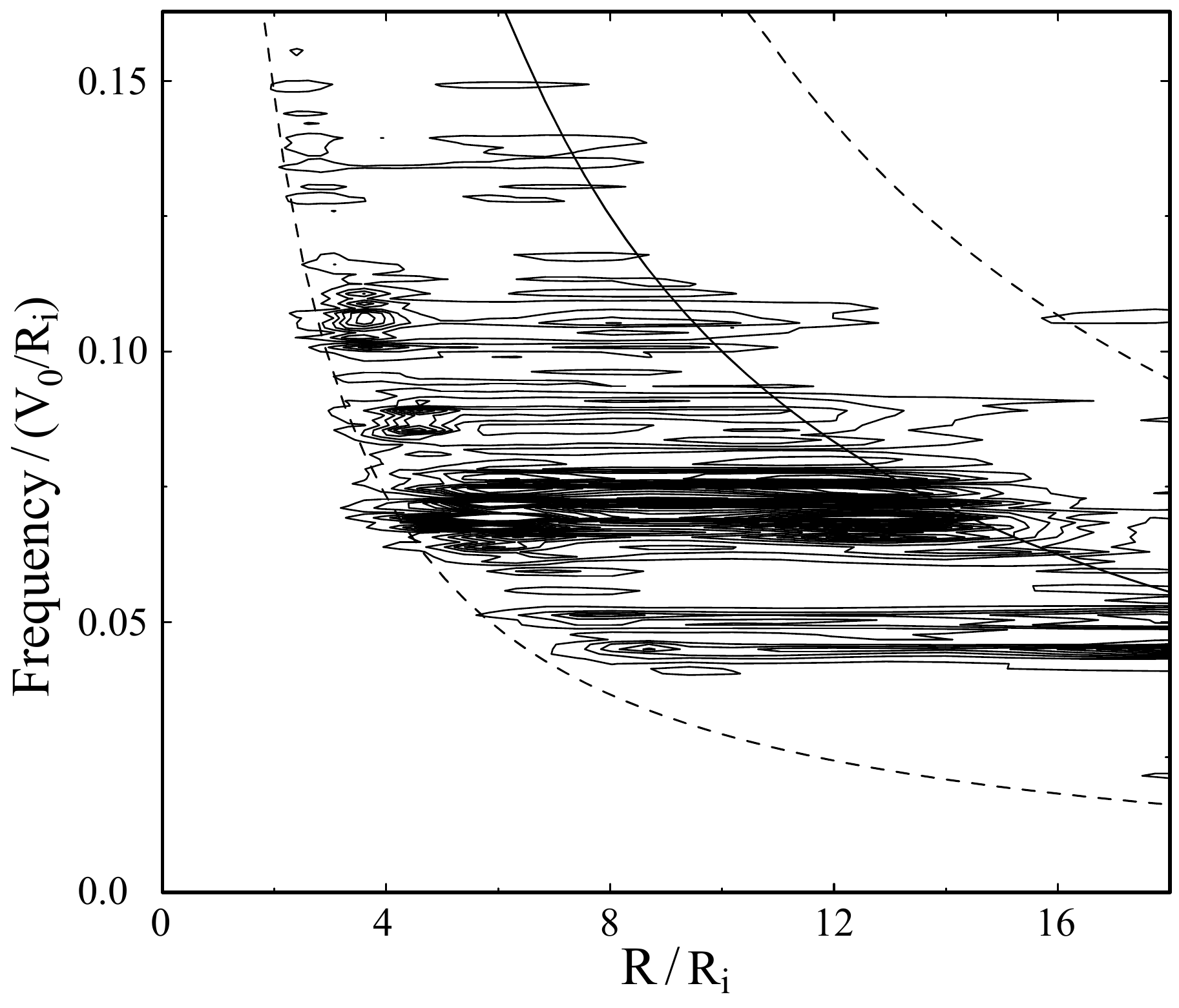}
\caption{The power spectrum of $m=2$ density variations in simulation
  UC.  The solid curve indicates the radius of corotation for the given
  frequency, while the dashed curves show the radii of the Lindblad
  resonances.}
\label{fig:run104spct}
\end{figure}

Fig.~\ref{fig:run104spct} illustrates the $m=2$ power spectrum of
disturbances.  Each horizontally extended peak indicates a spiral of a
particular pattern speed.  Some $20$ transient spirals of significant
amplitude occur during this period spanning a wide range of angular
frequencies and corotation radii.

\begin{figure}
\includegraphics[width=84mm]{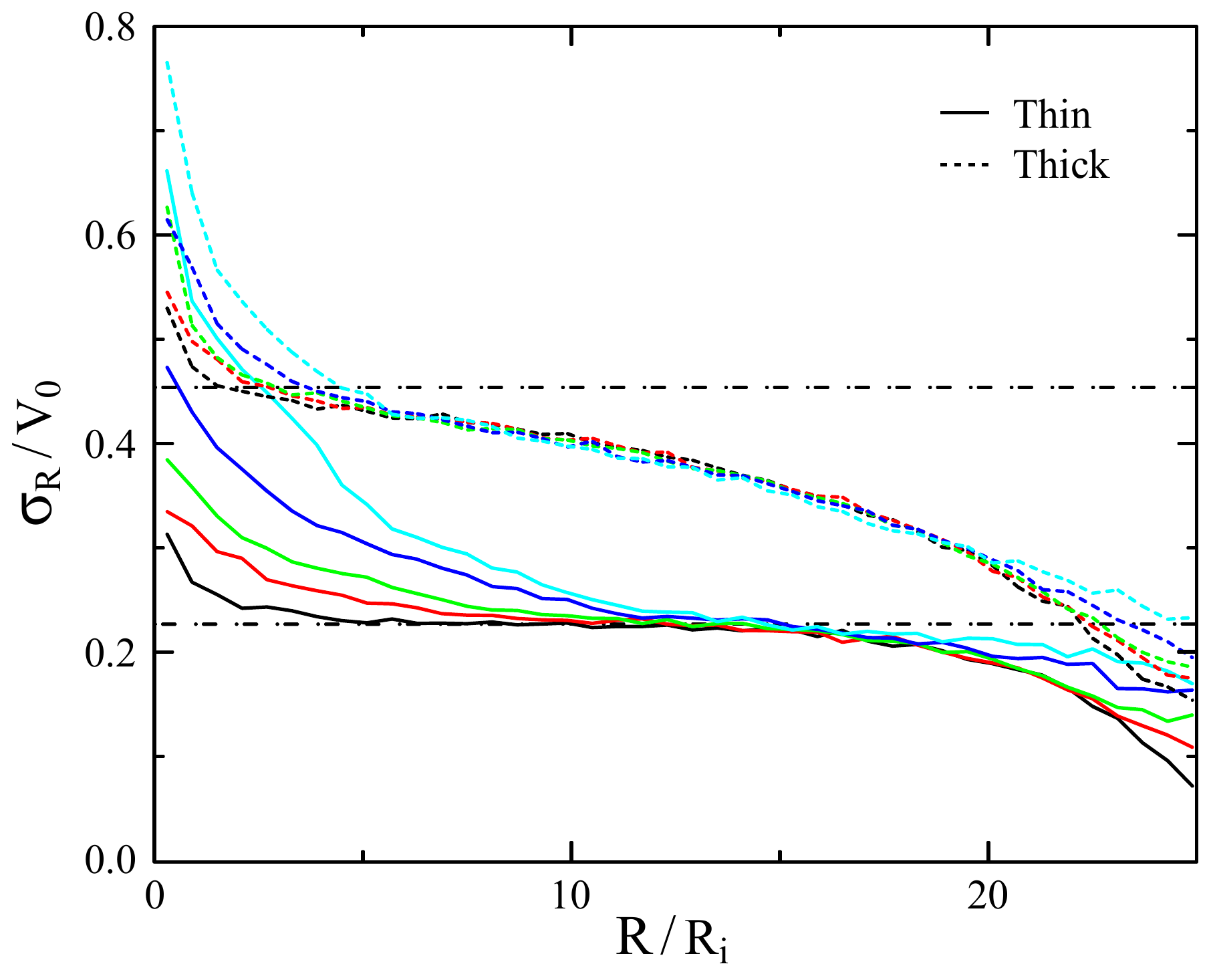}
\caption{Radial variations of $\sigma_R$ in simulation UC for the thin
  (solid) and thick (dashed) discs.  The curves are drawn at equal time
  intervals and $\sigma_R$ rises monotonically in the inner and outer
  discs.}
\label{fig:run104sigmar}
\end{figure}

Since there are numerous disturbances with well scattered Lindblad
resonances, random motion rises generally over the disc, as reported
in Fig.~\ref{fig:run104sigmar}, in contrast to Fig.~\ref{fig:run94sigmar}
which shows the localized heating at the ILR of the single spiral case.

\begin{figure}
\includegraphics[width=84mm]{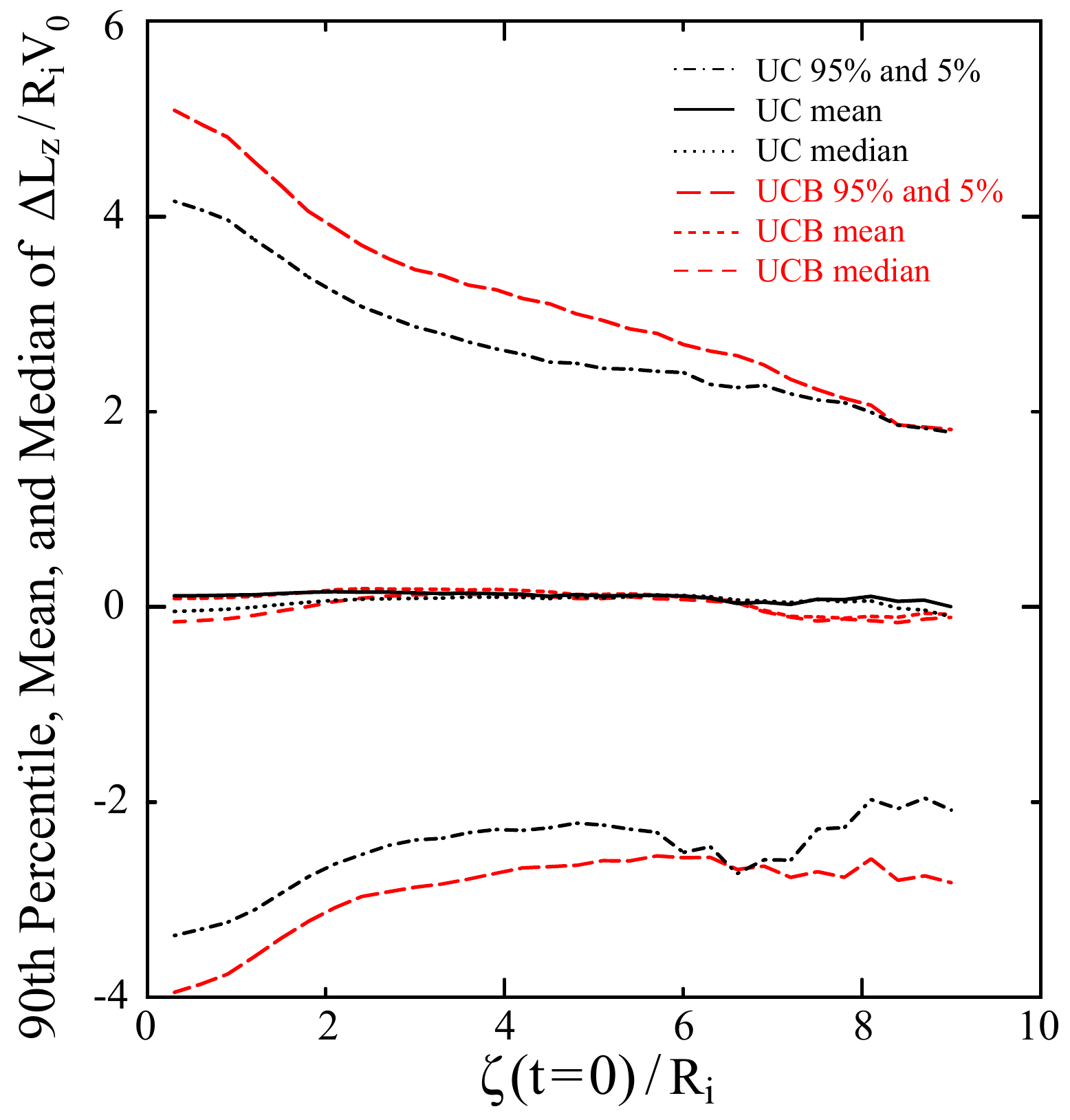}
\caption{Same as Fig.~\ref{fig:run949799101lchmx} for simulations UC
  (dot-dashed black) and UCB1 (long-dashed red).  The solid black and
  short-dashed red curves show the average angular momentum changes
  for UC and UCB1 respectively, and the dotted black and medium-dashed
  red ones show the median changes.}
\label{fig:run104lchmx}
\end{figure}

The larger changes in $L_z$ than those for the single spiral case are
evident from the dot-dashed black curves of Fig.~\ref{fig:run104lchmx},
which show the 5th and 95th percentiles of $\Delta L_z$ versus initial
notional vertical amplitude $\zeta$.  Although the shapes of these
curves are similar to those in Fig.~\ref{fig:run949799101lchmx}, they
are asymmetric about zero indicating that gains are larger than
losses.

\begin{figure}
\includegraphics[width=80mm]{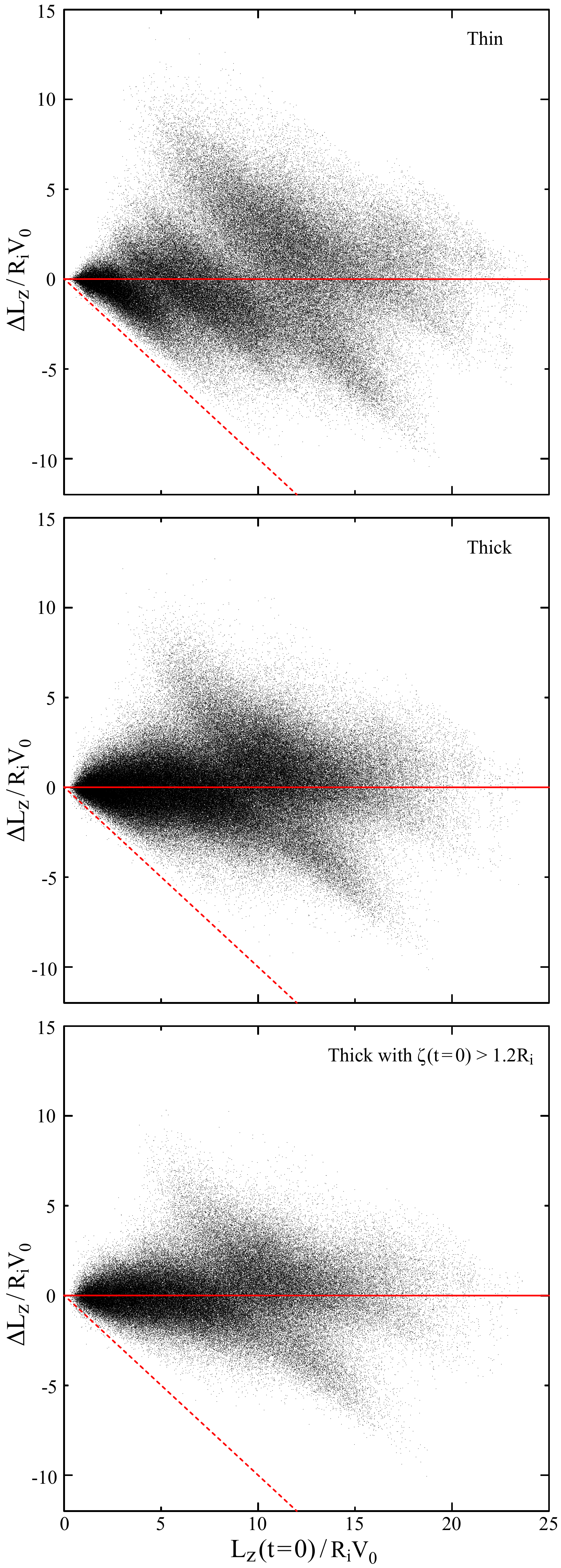}
\caption{Same as Fig.~\ref{fig:run94lch} for simulation UC.}
\label{fig:run104lch}
\end{figure}

\begin{figure}
\includegraphics[width=84mm]{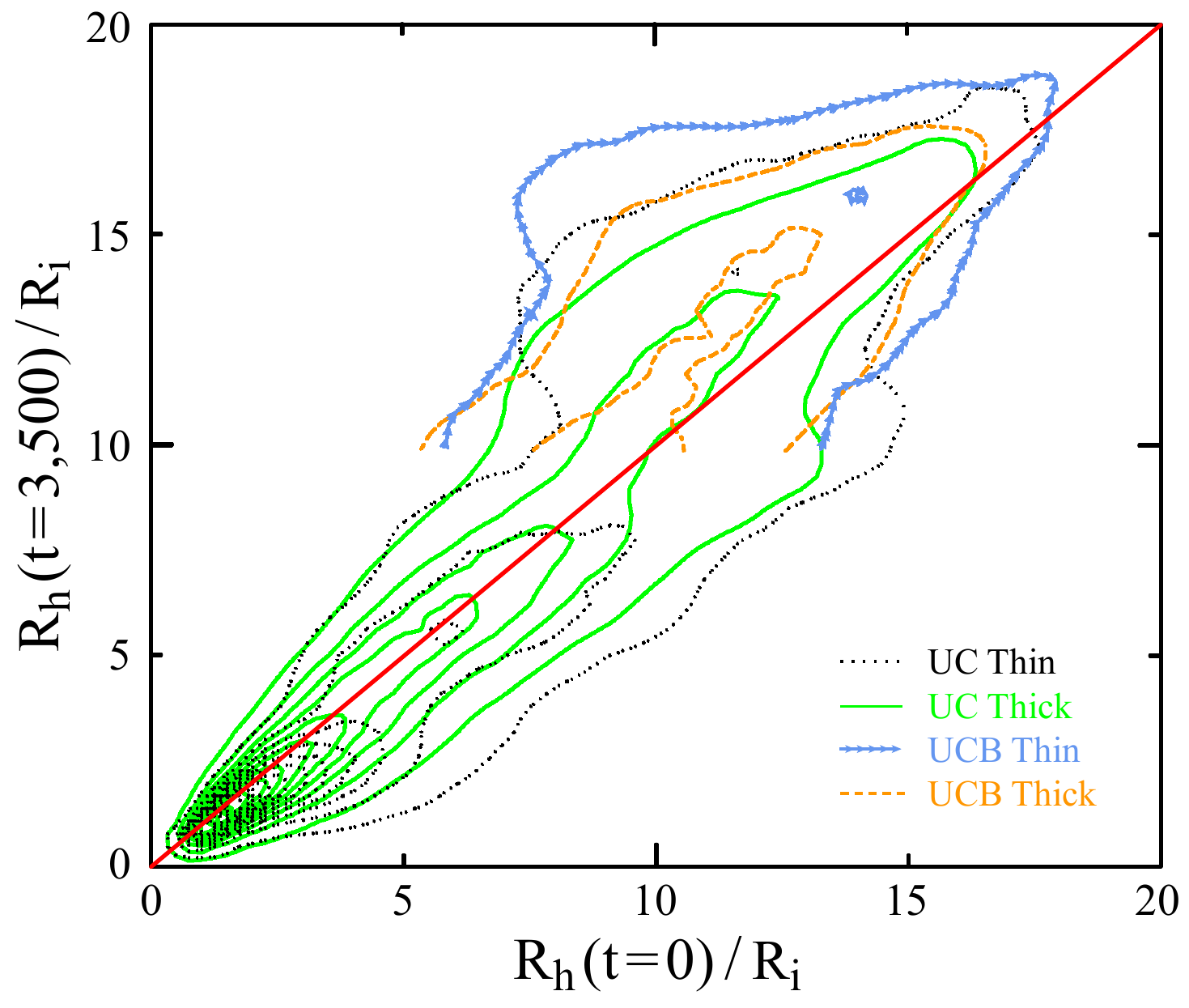}
\caption{The distributions of final home radii versus initial home
  radii for all the particles in simulation UC.  The particle density
  in this plane is estimated using an adaptive kernel, the dotted black
  and solid green contours represent the thin and thick discs respectively,
  and the red line shows zero change in $R_h$.  Contour levels are
  chosen every $10\%$ of the thick disc's maximum value from $5\%$ to
  $95\%$ for both thin and thick discs.  The arrow light blue and dashed
  orange contours represent the same for the thin and thick discs of UCB1
  respectively.  The region inside the bar ($R_h(t=3500) < 10R_i$) is
  omitted.}
\label{fig:run104rhfrhi}
\end{figure}

The angular momentum changes illustrated in Fig.~\ref{fig:run104lch}
reflect mostly the effects of the latest spirals that developed in the
simulation in each $L_z(t=0)$ region.  Contours of changes to the
distribution of home radii for the particles (dotted black and solid
green in Fig.~\ref{fig:run104rhfrhi}) reveal that particles from all
initial radii can move to new home radii, but that the changes are
greatest around the mid range of initial radii where corotation
resonances are more likely.  Some $0.003\%$ of the particles in the
thin disc and $0.03\%$ in the thick end up on retrograde orbits.

Taken together with the results from the single spiral simulations
described above, we again find that the changes in angular momenta and
home radii are almost as great in the thick disc as in the thin.  That
is, radial migration is weakened only slightly by disc thickness.

\begin{figure}
\includegraphics[width=80mm]{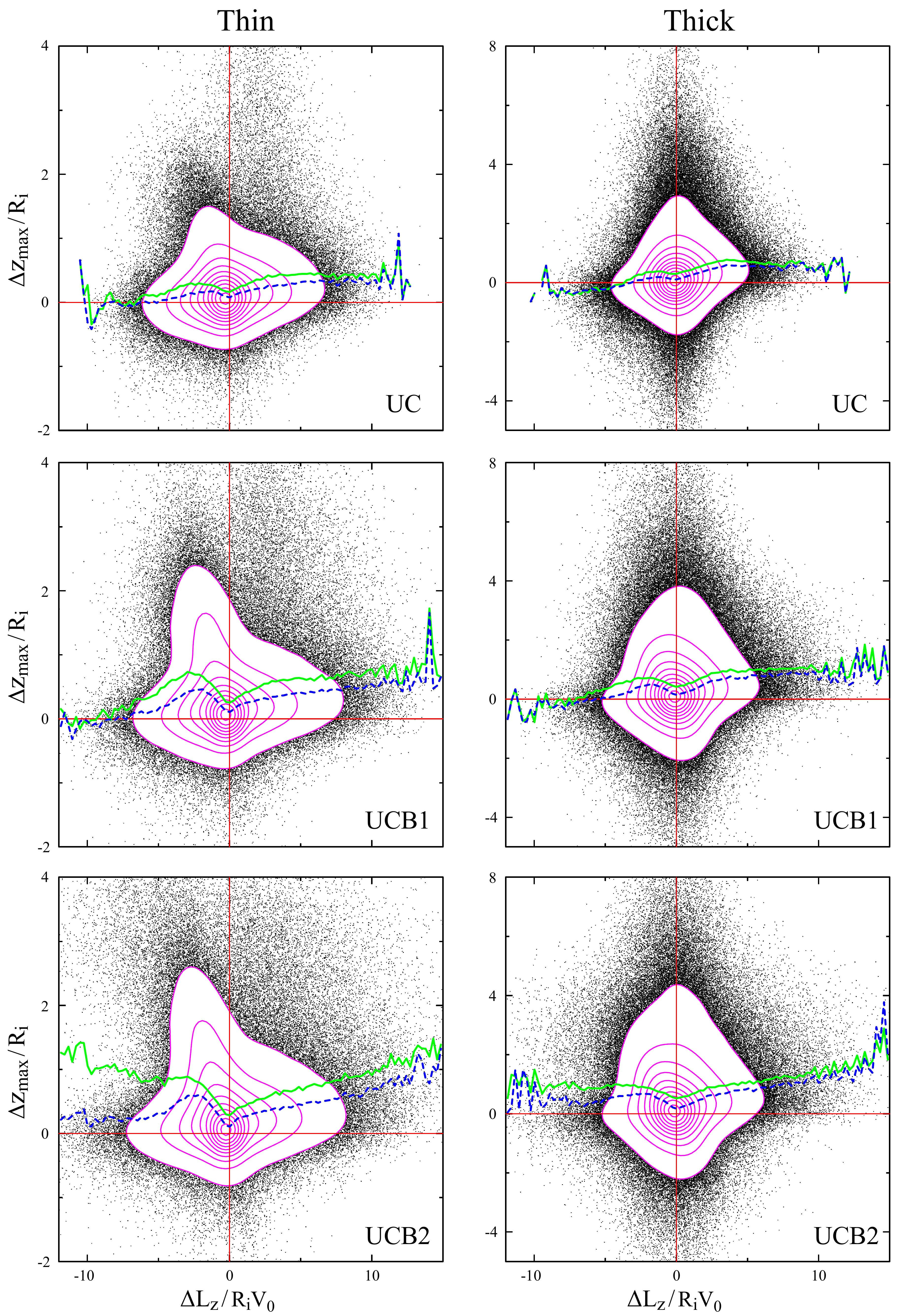}
\caption{Same as Fig.~\ref{fig:run94zmaxchlch} for simulations UC (top row),
  UCB1 (middle row), and UCB2 (bottom row).  The left column shows
  $\Delta z_{\rm max}$ as a function of $\Delta L_z$ for the thin discs and
  the right for the thick.}
\label{fig:run104zmaxchlch}
\end{figure}

The top row of Fig.~\ref{fig:run104zmaxchlch} shows changes in the UC
particles' maximum vertical excursions versus changes in their angular
momenta.  We find that, on average, $z_{\rm max}$ increases except for
the greatest losses in angular momentum.  This is different from the
single spiral case, in which the contours and the mean and median
curves are centred on the origin (Fig.~\ref{fig:run94zmaxchlch}).
This overall extra increase in vertical amplitude probably comes from
the net heating by vertical resonances that the multiple transient
spirals induce by the end of the simulation.  Nevertheless, the trend
of the mean and median with $\Delta L_z$ remains as in M2.  An
exception is again the group of points along the slope of $-1$ in the
second quadrant.  It is much more pronounced for the thin disc than in
M2.  The particles contributing to this feature are still initially
from the inner region of the disc, but this region is more extended in
UC since the inner vertical resonances occur at various radii for the
numerous transient spirals.  As in M2, changes in $z_{\rm max}$ remain
about twice as great for the thick disc particles as for those of the
thin for the same $\Delta L_z$.

Although the scale height of outward migrating particles does increase
somewhat, as expected, the changes are not substantial enough to cause
the thickness of outward migrating thin-disc particles to approach the
scale height of the thick disk.  The changes in the vertical motion of
inward migrating particles are more minor than those for outward
migrators.  Thus we do not observe much of a tendency in our models
for evolution to cause a significant degree of blurring between the
separate populations.

\subsection{Multiple spirals with a bar}
We here study radial migration in two simulations, UCB1 and UCB2, that
formed bars at an early stage of their evolution and compare them with
simulation UC that did not form a bar for a long period.  As noted in
the introduction, it has long been known that bar formation causes
some of the largest changes to the distribution of angular momentum
within a disc.  Here our focus is on the consequences of continued
transient spiral behaviour in the outer disc long after the bar
formed, which has not previously received much attention, as far as we
are aware.

The red curve of Fig.~\ref{fig:run104ampl} shows the evolution of
$A_2/A_0$ in UCB1.  The spirals become significant around the time
$\sim 1\,000R_i/V_0$, the bar forms around $\sim 2\,000R_i/V_0$, and
we stop the simulation at the same final time $t = 3\,500R_i/V_0$ as
UC.  Thus, the time interval of significant scattering is roughly
$2\,500R_i/V_0$ in length with a bar being present for the last $\sim
1\,500R_i/V_0$, which correspond to $7.5\;$Gyr and $4.5\;$Gyr
respectively.  With our suggested scaling, the bar has a pattern speed
of $31.8\;$km~s$^{-1}$~kpc$^{-1}$ and corotation $R_c \sim 7.5\;$kpc.
(This scaling makes the bar substantially larger than that in the
Milky Way.)

The long-dashed red curves in Fig.~\ref{fig:run104lchmx} show that the
bar enhances the changes in $L_z$ somewhat over those that arise due
to spirals alone.  Note that we measure the instantaneous value of
$L_z$ of each particle and it should be borne in mind that it changes
continuously in the strongly non-axisymmetric potential of this model,
especially so for particles in or near the bar.

For this reason, we cannot extend the light blue (marked with arrow
heads) and dashed orange contours in Fig.~\ref{fig:run104rhfrhi} to
small home radii at the later time in the strongly non-axisymmetric
potential of the bar.  However, a clear asymmetry can be seen; the
distribution is biased above the zero change red line, indicating a
systematic outward migration in the outer disc.  A similar asymmetry
can be seen without a bar in UC (dotted black and solid green
contours), but the formation of the bar makes it more pronounced.
Since total angular momentum is conserved in these simulations, there
is a corresponding inward migration in the inner regions (see
Fig.~\ref{fig:run94lch}).

\begin{figure}
\includegraphics[width=84mm]{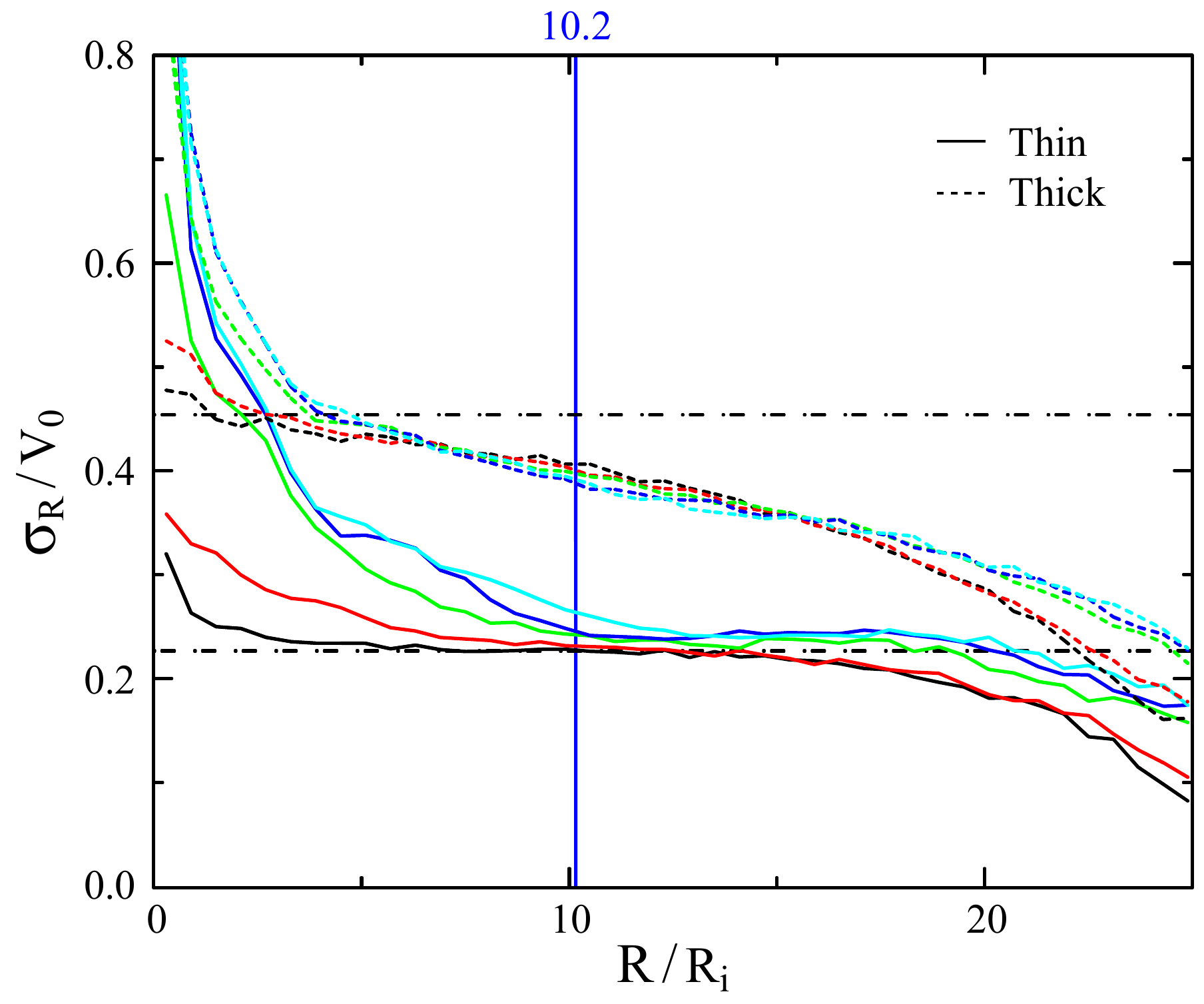}
\caption{Same as Fig.~\ref{fig:run104sigmar} for UCB1.}
\label{fig:run104Bsigmar}
\end{figure}

Aside from the bar region, where systematic non-circular streaming
biases the rms radial velocities, Fig.~\ref{fig:run104Bsigmar} shows
that the bar does not appear to cause much extra heating over that in
the unbarred case.

\begin{figure}
\includegraphics[width=80mm]{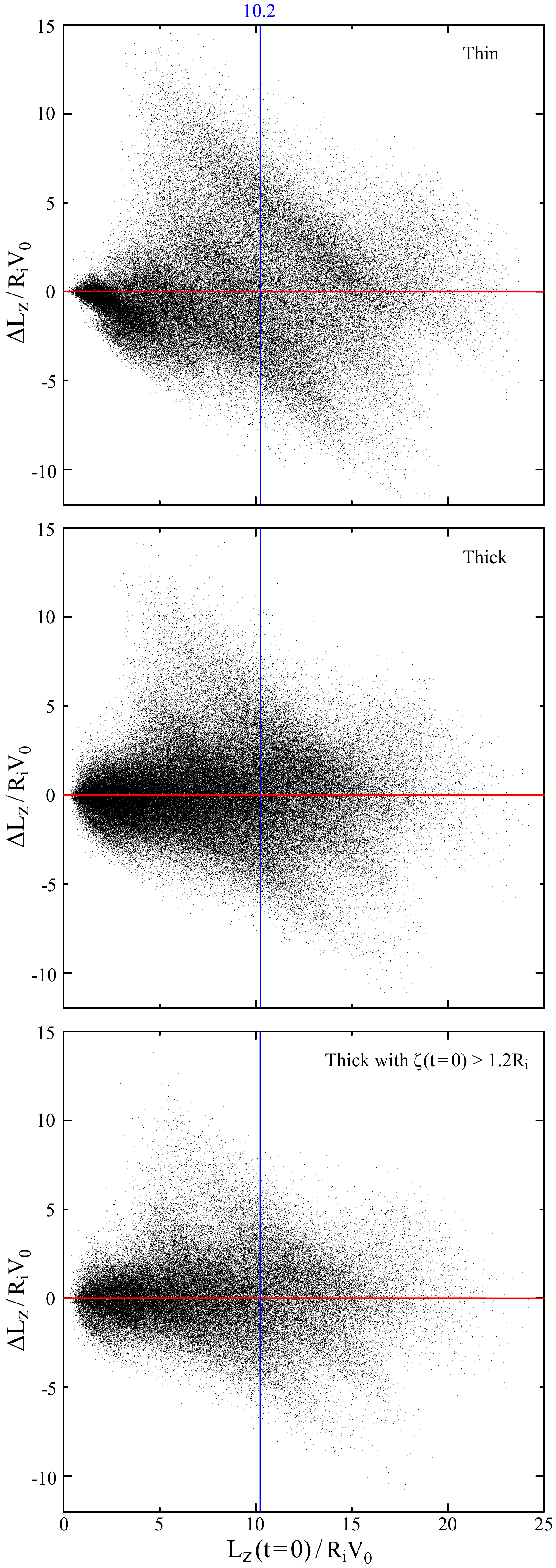}
\caption{Same as Fig.~\ref{fig:run104lch} for simulation UCB1.  The
  vertical blue line marks the approximate corotation resonance of
  the bar.}
\label{fig:run104Blch}
\end{figure}

Fig.~\ref{fig:run104Blch} again illustrates that changes in $L_z$ in
the outer disc are more substantial in this barred model than in the
unbarred case (Fig.~\ref{fig:run104lch}), but the larger spread in
$\Delta L_z$ in the inner disc is partly an artifact of using the
instantaneous values of $L_z$ at later times.  Because we use the
instantaneous $L_z$ in the barred potential in this Figure, a particle
in the region where $\Delta L_z < -L_z$ need not necessarily have been
changed to a fully retrograde orbit.

The asymmetries in Figs.~\ref{fig:run104rhfrhi} \&
\ref{fig:run104Blch} about the red lines of zero change are not caused
by bar-formation alone, as the distributions (not shown) are
approximately symmetric immediately after this event.  Rather, we find
they appear to be caused by multiple migrations of the same particles
by separate events.

Our second barred simulation, UCB2, again differs from UCB1 and UC
only by the initial random seed.  It formed a stronger bar, but a
little later than in UCB1, as shown by the green curve of
Fig.~\ref{fig:run104ampl}.  Significant scattering occurs for the last
$\sim 1\,800R_i/V_0$ time units ($5.4\;$Gyr) of which the bar is present
for the last $\sim 1\,100R_i/V_0$ ($3.3\;$Gyr).  The pattern speed
($33.6\;$km~s$^{-1}$~kpc$^{-1}$) and corotation radius ($\sim
7.3\;$kpc) for our adopted scaling, are similar to the values in UCB1.

We find the extent of radial migration (bottom two rows of
Table~\ref{tab:rmslch}) is further increased by the stronger bar, but
not by much.  Variants of Figs.~\ref{fig:run104Blch} \&
\ref{fig:run104rhfrhi} (not shown) are qualitatively similar with
slightly larger changes, but the outer disc is again dominated by
scattering due to the latest few spirals.

In both these models, therefore, we see that the formation of a bar
does indeed increase the net angular momentum changes.  However the
overall behaviour is similar to that of scattering by transient
spirals without a bar.  This is different from the effect found by
\citet{BC11}, in which essentially all the angular momentum changes
occurred during bar formation.

The lower two rows of Fig.~\ref{fig:run104zmaxchlch} show $\Delta
z_{\rm max}$ versus $\Delta L_z$ for UCB1 and UCB2 respectively.  The
presence of a bar yields a much denser and more extended feature in
the second quadrant, which is also visible in for thick disc.  For
UCB2, its contribution is so great that the mean $\Delta z_{\rm max}$
keeps increasing with greater angular momentum loss.  This extra
apparent vertical heating is probably caused by the buckling of the
bar.  Unlike for M2, UC, and UCB1, we find that for positive $\Delta
L_z$ in UCB2, the mean and median curves keep rising for larger
$\Delta L_z$.

\section{A Conserved Quantity?}
Radial migration results from angular momentum changes near
corotation, which we have shown to be somewhat weakened by increased
vertical motion.  \citet{S09a}, in their model of radial mixing in
thin and thick discs, assumed that vertical and radial motions are
decoupled and that vertical energy is conserved as stars migrate
radially.  We here try to identify a conserved quantity that can be
used to predict vertical motion when particles suffer large changes in
$L_z$, by comparing various measures of vertical amplitude at the
initial and final times in simulations with a single spiral.  Note
that the ``initial'' value of each quantity we discuss in this section
is measured at $t=64R_i/V_0$, which avoids possible effects of the
settling of the model from its mild initial imbalance.  Also, all
quantities are computed in an azimuthally averaged potential, to
eliminate variations with spiral phase, which remain significant at
the final time.

We focus on particles whose initial home radii lie in an annulus of
width $6.0R_i$ centred at corotation of the spiral in M2, and measure
quantities for only one particle per quiet start ring, meaning one in
every twelve.  This results in $\sim 40\,000$ particles from the thin
disc and $\sim 73\,000$ from the thick.  Although the entire thick
disc is represented by just 50\% more particles than is the thin, the
effects of the inner and outer tapers, together with the groove in the
thin disc cause the number of particles in the range
$4.0R_i \le R_h(t=64) \le 10.0R_i$ to be some 82\% larger for the thick
disc than for the thin.  While we endeavour to measure each quantity
in this section for the same set of particles, from both simulations M2
and T, we have been able to estimate some quantities for only a subset
of these particles, as noted below.

\begin{figure}
\includegraphics[width=80mm]{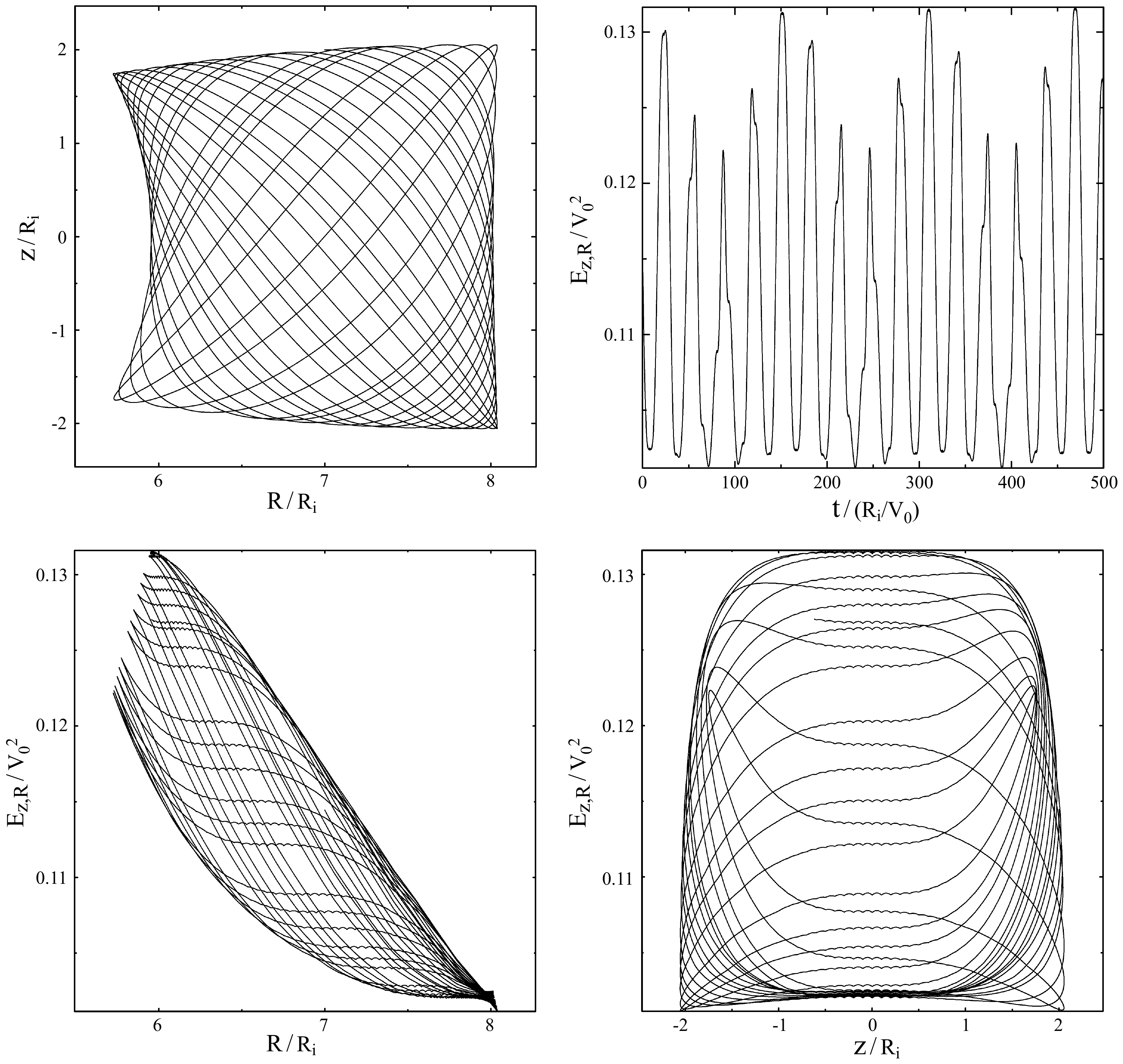}
\caption{The top left panel shows a typical thick disc orbit in the
  meridional plane, and the other three panels show how $E_{z,R}$ varies
  with time (top right), radius (bottom left), and $z$-height (bottom
  right).  The particle, which has a home radius of $6.5R_i$, a total
  energy of $-2.0V_0^2$, a midplane eccentricity of $0.15$, and a vertical
  excursion of $2.1R_i$, is integrated in the frozen initial axisymmetric
  potential of simulation M2.}
\label{fig:run94orbvenrg}
\end{figure}

\subsection{Vertical energies}
We have tried a number of different way to estimate the energy of
vertical motion.  A simple definition might be
\begin{equation}
E_{z,R} = \frac{1}{2}v_z^2 + \Phi(R,z) - \Phi(R,0),
\label{Ezr}
\end{equation}
with $R$ being the instantaneous radius of the particle.  However,
Fig.~\ref{fig:run94orbvenrg} shows that $E_{z,R}$ defined this way
varies by some 30\% as a particle oscillates in both radius and
vertically in a static axisymmetric potential.  The orbit of this
particle is multiply periodic and clearly respects three integrals (as
we will confirm below), in common with many orbits in axisymmetric
potentials (BT08, section 3.2).  Although such orbits can be described
by action-angle variables, which imply three decoupled oscillations,
the {\em energy} of vertical motion is clearly not decoupled from that
of the radial part of the motion.  In particular, the bottom left
panel shows that $E_{z,R}$ varies systematically with radius, which
reflects in part the weakening of the vertical restoring force with
the outwardly declining surface density of the disc.  This behaviour
considerably complicates our attempts to compare the vertical energies
at two different times in the same simulation.

The epicycle approximation (BT08, p.~164) holds for stars whose orbits
depart only slightly from circular motion in the midplane.  In this
approximation, when a star pursues a near-circular orbit near the
mid-plane of an axisymmetric potential, the vertical and radial parts
of the motion are separate, decoupled oscillations, and the vertical
energy is $E_{z,\rm epi} \;(=Z$ eq.~\ref{Ez}) is constant.  However,
the epicycle approximation is a poor description of the motion of most
particles, for which the radial and vertical oscillations are neither
harmonic nor are the energies of the two oscillations decoupled.

Table~\ref{tab:EzJzBiweight} gives the rms values of $[Y(t_{\rm
    final}) - Y(t_{\rm initial})] / Y(t_{\rm initial})$ for particles
in both the thin and the thick discs of simulations M2 and
T.\footnote{For all quantities in this table, we use the biweight
  estimator of the standard deviation \citep{Beers}, which ignores
  heavy tails.  We also discard a few particles with initial values
  $<10^{-4}$ in order to avoid excessive amplification of errors
  caused by small denominators.}  Here $Y$ represents one of several
possible vertical integrals.  The first rows give the fractional
changes in the epicycyclic approximation, $Y = E_{z,\rm epi}$, that
are substantial.  Furthermore, they are almost as large in simulation
T, which was constrained to remain axisymmetric, as those in M2 in
which substantial radial migration occurred, suggesting that the
changes are mostly caused by the inadequacy of the approximation.

\begin{table}
\caption{Biweight standard deviation of fractional changes in various
  estimates of vertical energy and action.}
\label{tab:EzJzBiweight}
\begin{tabular}{@{}llccc}
\hline
Simulation & Disc & Variable $Y$ & \multicolumn{2}{c}{$\sigma(\Delta Y/Y_{\rm initial})$} \\
\hline
M2 & Thin & $E_{z,\rm epi}$ & 30.7\% & 28.9\% \\
   & & $E_{z,R}$ & 28.3\% & 27.2\% \\
   & & $E_{z,R_h}$ & 25.0\% & 26.1\% \\
   & & $\langle E_{z,R}\rangle$ & & 23.6\% \\
   & & $J_{z,\rm epi}$ & 22.7\% & 20.7\% \\
   & & $J_z|_{R,\phi}$ & 16.5\% & 16.0\% \\
   & & $J_z$ & & 15.6\% \\
   & Thick & $E_{z,\rm epi}$ & 59.4\% & 39.4\% \\
   & & $E_{z,R}$ & 47.6\% & 31.5\% \\
   & & $E_{z,R_h}$ & 34.9\% & 26.0\% \\
   & & $\langle E_{z,R}\rangle$ & & 22.3\% \\
   & & $J_{z,\rm epi}$ & 55.7\% & 34.9\% \\
   & & $J_z|_{R,\phi}$ & 35.0\% & 19.7\% \\
   & & $J_z$ & & 15.4\% \\
T & Thin & $E_{z,\rm epi}$ & 22.5\% & 19.5\% \\
  & & $E_{z,R}$ & 17.9\% & 15.4\% \\
  & & $E_{z,R_h}$ & 10.0\% & 8.8\% \\
  & & $\langle E_{z,R}\rangle$ & & 5.8\% \\
  & & $J_{z,\rm epi}$ & 16.5\% & 13.8\% \\
  & & $J_z|_{R,\phi}$ & 9.4\% & 8.2\% \\
  & & $J_z$ & & 6.1\% \\
  & Thick & $E_{z,\rm epi}$ & 53.6\% & 34.3\% \\
  & & $E_{z,R}$ & 40.8\% & 22.8\% \\
  & & $E_{z,R_h}$ & 23.3\% & 11.4\% \\
  & & $\langle E_{z,R}\rangle$ & & 3.0\% \\
  & & $J_{z,\rm epi}$ & 47.5\% & 27.8\% \\
  & & $J_z|_{R,\phi}$ & 24.8\% & 11.0\% \\
  & & $J_z$ & & 3.4\% \\
\hline
\end{tabular}

\medskip
The biweight estimated standard deviation of the fractional change in
variable $Y$ listed in the third column for both the thin and thick
discs in simulations M2 and T.  We give two values for the standard
deviation in some rows: the value in the fourth column is measured
from all the particles, that in the fifth column is from only the 48\%
(in M2) of particles for which $\Delta J_z$ is calculable.
\end{table}

The second rows in Table~\ref{tab:EzJzBiweight} show the rms
fractional changes in the instantaneous values of $E_{z,R}$.  While
these values are slightly smaller than for the epicyclic estimate,
they are again large for the thin disc and still greater for the
thick.  This is hardly surprising, as the particles in the simulation
have random orbit phases at the two measured times.

Particles on eccentric orbits generally spend more time at radii $R >
R_h$ than inside this radius, which biases the instantaneous measure
to a lower value, as shown in the third panel of
Fig.~\ref{fig:run94orbvenrg}.  To eliminate this bias, and to reduce
the random variations, we evaluate $E_{z,R_h}$ from eq.~(\ref{Ezr}) at
the home radius of each particle, which requires us to integrate the
motion of each particle in the frozen potential of the appropriate
time until the particle reaches its home radius.  A tiny fraction,
$\sim 500$ of the $\sim 113\,000$ sample particles, never cross
$R=R_h$ at the initial time and $\sim 200$ more at the final time;
these particles rise to large heights above the mid-plane and have
meridional orbits resembling that shown in the top right of
Fig.~\ref{fig:run94vactorbts}, but appear to be confined to $R>R_h$.
The fact that this is possible seems consistent with the description
of vertical oscillations developed by \citet{S12}.  The third rows of
Table~\ref{tab:EzJzBiweight} give the fractional rms changes in
$E_{z,R_h}$ for all the remaining particles; the changes in simulation
M2 are a little smaller than those of the instantaneous values and
significantly so in simulation T.

Since $E_{z,R}$ is multiply periodic (Fig.~\ref{fig:run94orbvenrg}),
we have also estimated an orbit-averaged value, $\langle
E_{z,R}\rangle$, by integrating the motion for many periods until the
time average changed by $<0.1\%$ when the integration is extended for
an additional radial period.  The fourth rows of
Table~\ref{tab:EzJzBiweight} give the rms changes in $\langle
E_{z,R}\rangle$, which are still considerable.  Note that we did not
compute this time-consuming estimate for all the particles, but for
only the subset used for other values in the fifth column of this
table.  The choice of this subset is described below.

Generally, we find that none of these estimates of the vertical energy
is even approximately conserved, except for the orbit-averaged energy
when the simulation is constrained to remain axisymmetric (simulation
T).  In this case, the potential at the two times differs slightly due
to radial variations in the mass distribution -- values of this
orbit-averaged quantity in an unchanged potential would be independent
of the moment at which the integration begins.

\begin{figure}
\includegraphics[width=84mm]{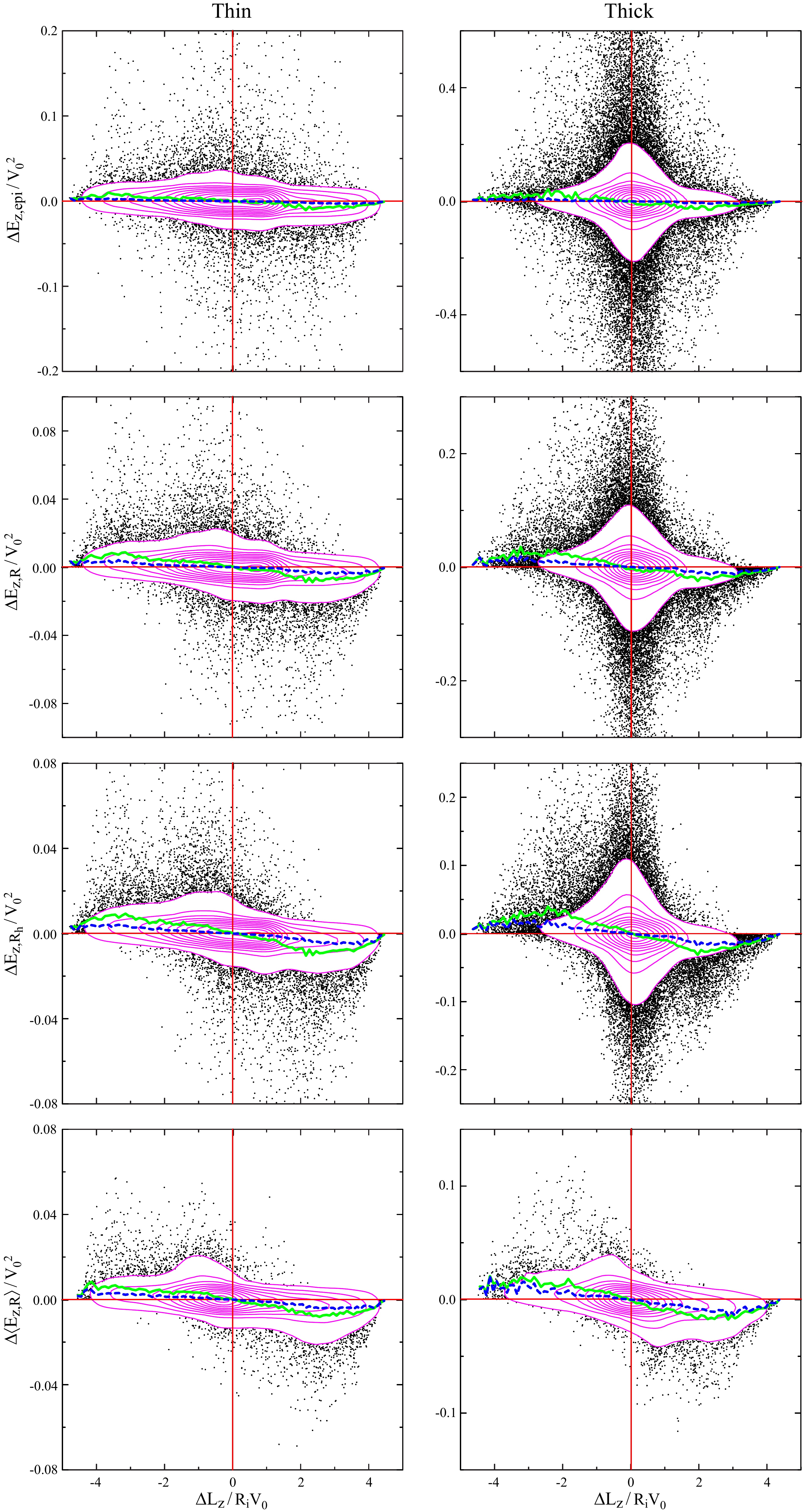}
\caption{Changes in $E_{z,\rm epi}$ (first row), $E_{z,R}$ (second
  row), $E_{z,R_h}$ (third row), and $\langle E_{z,R}\rangle$ (fourth
  row) versus $\Delta L_z$ for the thin (left column) and thick (right
  column) discs of simulation M2.  Just as in Fig.~\ref{fig:run94zmaxchlch},
  the horizontal and vertical red lines show zero changes, the linearly
  spaced magenta contours show number density, and the bold solid green
  and dashed blue curves show the mean and median changes in each
  ordinate respectively.  Note that the vertical scales differ in each plot.}
\label{fig:run94ezchlch}
\end{figure}

In all cases, changes are larger in simulation M2 where significant
radial migration occurs.  Fig.~\ref{fig:run94ezchlch} illustrates that
the changes in the estimated vertical energy correlate with $\Delta
L_z$.  The excesses of particles in the second and fourth quadrants
indicate that changes have a tendency to be negative for outwards
migrating particles and positive for inwards migrating particles.  The
significance of this trend is discussed below.

\subsection{Vertical actions}
The various actions of a regular orbit in a steady potential are
defined to be $(2\pi)^{-1}$ times the appropriate cross-sectional area
of the orbit torus (BT08, pp.~211--215).  One advantage of actions is
that they are the conserved quantities of an orbit when the potential
changes slowly -- \ie\ they are adiabatic invariants under conditions
that are defined more carefully in BT08 (pp.~237--238).

\begin{figure}
\includegraphics[width=84mm]{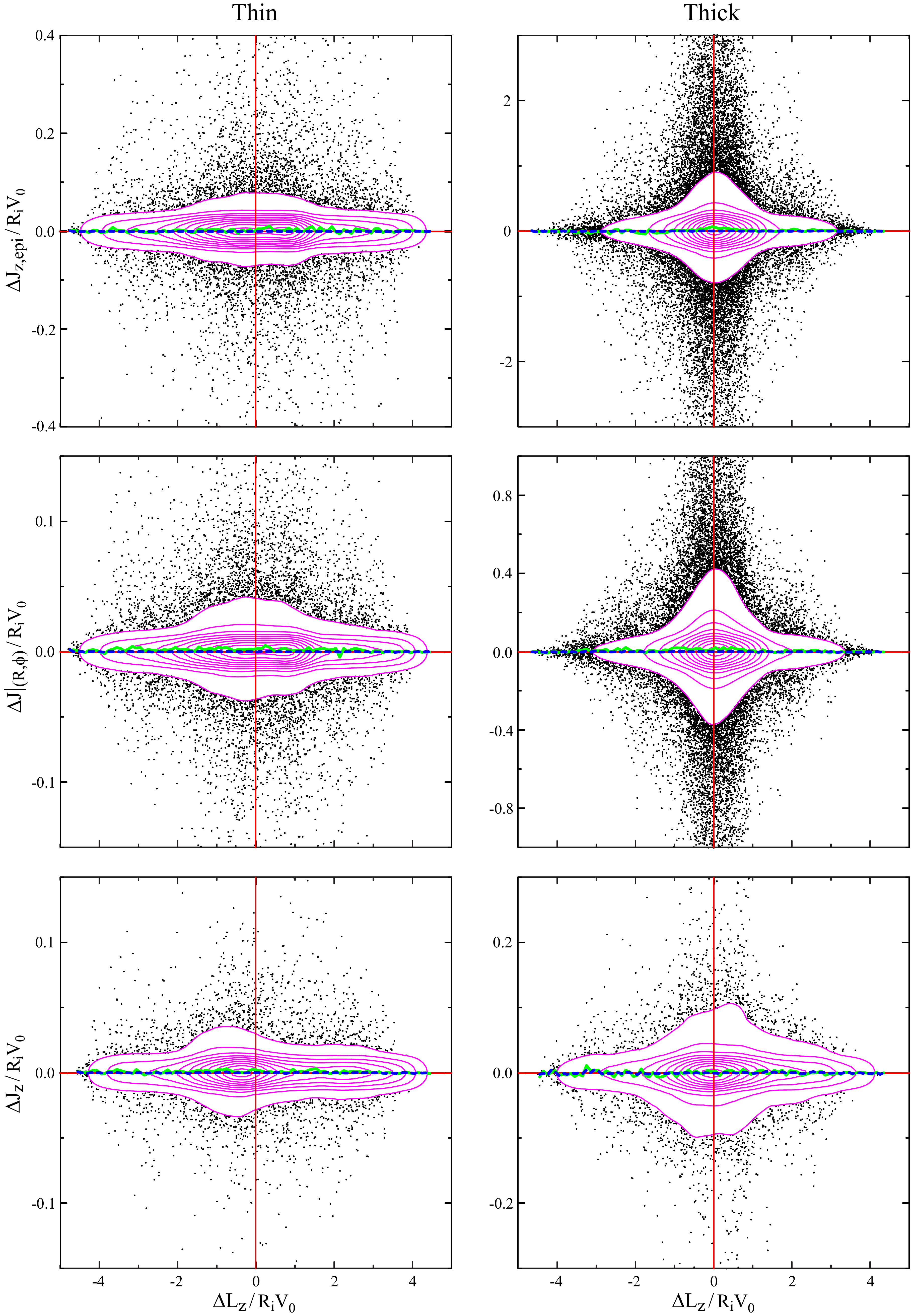}
\caption{ Same as Fig.~\ref{fig:run94ezchlch} for changes in $J_{z,\rm
    epi}$ (first row), $J_z|_{R,\phi}$ (second row), and $J_z$ (third
  row) of the particles in M2.}
\label{fig:run94jzchlch}
\end{figure}

The vertical action is defined as
\begin{equation}
J_z \equiv {1 \over 2\pi} \oint\dot{z}dz,
\label{vertact}
\end{equation}
where $z$ and $\dot z$ are measured in some suitable plane that
intersects the orbit torus.  Since $L_z \;(\equiv J_\phi)$ is
conserved in a steady axisymmetric potential, the orbit can be
followed in the meridional plane (BT08, p159) in which it oscillates
both radially and vertically, as illustrated in the first panel of
Fig.~\ref{fig:run94orbvenrg}.  To estimate $J_z$, we need to integrate
the orbit of a particle in a simulation from its current position in
the frozen, azimuthally averaged potential at that instant and
construct the $(z, \dot z)$ surface of section (SoS) as the particle
crosses $R_h$ with $\dot R>0$, say.  The integral in
eq.~(\ref{vertact}) is the area enclosed by an invariant curve in this
plane.

Before embarking on this elaborate procedure, we consider two possible
approximations.  When the epicycle approximation holds, the vertical
action is $J_{z,\rm epi} = E_{z,\rm epi}/\nu$ (BT08 p.~232).  The
fifth rows of Table~\ref{tab:EzJzBiweight} for each disc show that
changes in this quantity are large and again they are similar to those
in simulation T, confirming once again that the epicycle approximation
is inadequate.

Since most orbits reach beyond the harmonic region ($|z| \la 0.4R_i$)
of the vertical potential, an improved local approximation is to
calculate
\begin{equation}
J_z|_{R,\phi} = \left. (2\pi)^{-1} \oint\dot{z}dz \right|_{R,\phi}
\end{equation}
at the particle's fixed position in the disc. This local estimate
still ignores the particle's radial motion, but gives a useful
estimate for the average vertical action that is also used in analytic
disc modeling \citep{Binn10}.  The function $\dot z(z)$ at this fixed
point is simply determined by the vertical variation of
$\Phi(R,\phi,z)$, and the area is easily found.  We evaluate
$J_z|_{R,\phi}$ at the particle's instantaneous position at the
initial and final times using the corresponding azimuthally averaged
frozen potential.  The rms variation of $\Delta J_z|_{R,\phi}$ is
given in the sixth rows of Table~\ref{tab:EzJzBiweight} for each disc.
We find that $\Delta J_z|_{R,\phi}$ in the thin discs of both M2 and T
are small, suggesting that this estimate of vertical action is more
nearly conserved.  However, the changes for the thick discs are still
large, and we conclude that this local estimate is still too
approximate.

We therefore turn to an exact evaluation of eq.~(\ref{vertact}) using
the procedure described in the Appendix.  Unfortunately, we can
evaluate $\Delta J_z$ only if the consequents in the SoS form a simple
invariant curve that allows $J_z$ to be estimated at both times.  We
find that only 48\% of the $\sim 113\,000$ particles have closed,
concave invariant curves at the initial time and only 73\% of those
retain these properties at the final time.  Although the thick disc
contains more particles than the thin, we are able to calculate
$\Delta J_z$ for a smaller fraction: we succeed with $\sim 26\,000$ in
the thin disc but only $\sim 13\,000$ in the thick.

Column five of Table~\ref{tab:EzJzBiweight} gives the rms changes of
all estimates of vertical energy and action for only these particles.
While the fractional rms values of the different energy estimates
remain large, they are generally smaller than those for the all
particles given in the fourth column, but only slightly so for
$J_z|_{R,\phi}$.  Thus the orbits for which $\Delta J_z$ can be
computed are not a random subset, but are biased to those for which
energy changes are smaller on average.

\begin{figure}
\includegraphics[width=84mm]{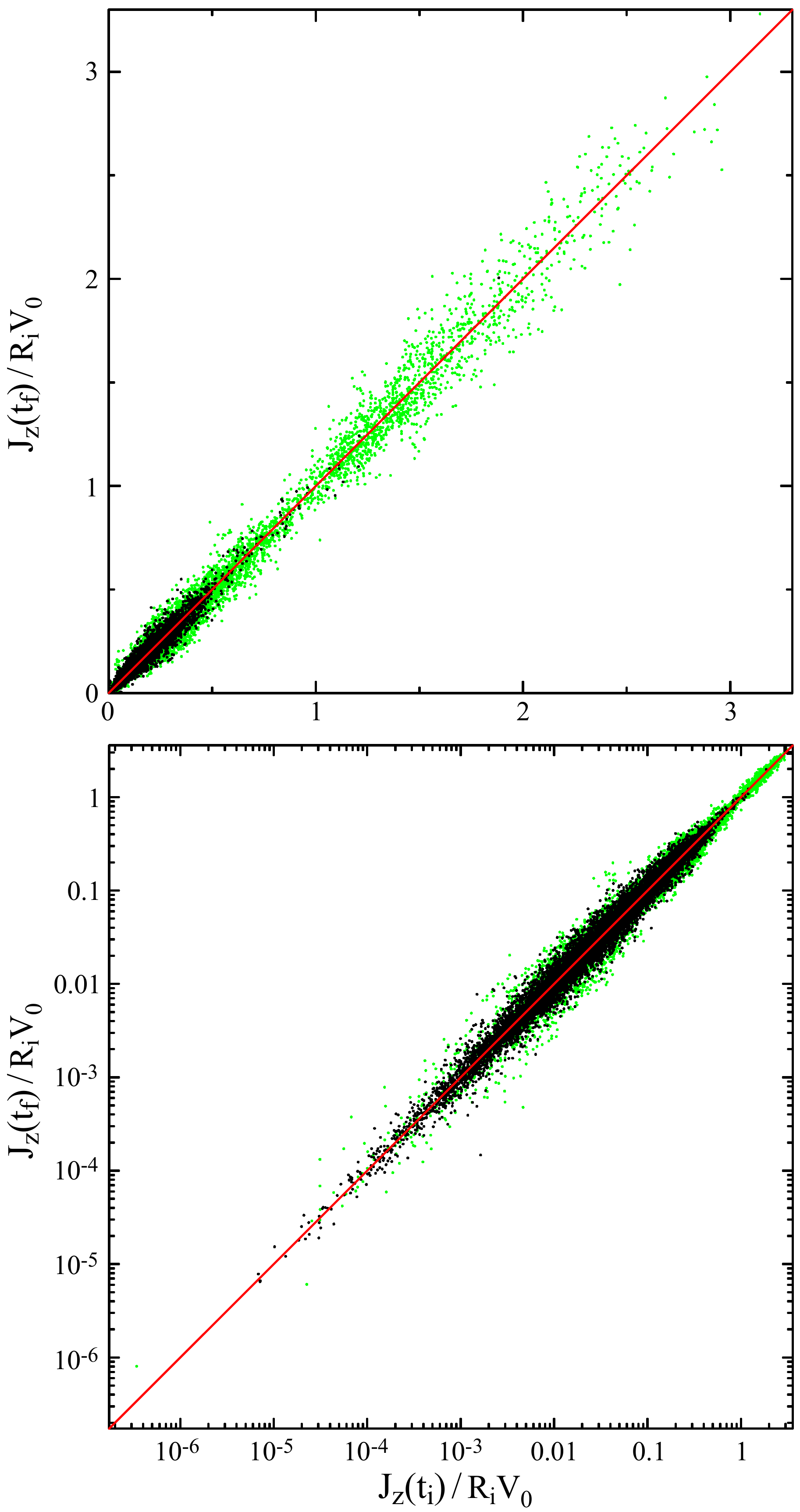}
\caption{Comparison of $J_z$ for $\sim 40\,000$ particles from
  simulation M2 at the initial and final times.  Values are calculated
  from eq.~(\ref{vertact}).  The top panel is on a linear scale, the
  bottom on a log scale to reveal the behaviour for very small actions.
  The red line has unit slope.  Thin disc particles are marked in black,
  thick disc in green.}
\label{fig:run94vact} 
\end{figure}

The seventh (final) rows of Table~\ref{tab:EzJzBiweight} for each disc
list the rms values of the fractional change in $\Delta J_z$.  They
are just a few percent for particles in simulation T, in which the
disc is constrained to remain axisymmetric, and are smaller than for
any other tabulated quantity in the discs perturbed by a spiral.  The
small scatter about the red line of unit slope in Fig.~\ref{fig:run94vact}
indicates that differences in the final and initial values of $J_z$ in
both discs for M2 are indeed small.

Fig.~\ref{fig:run94jzchlch} shows that changes in all three estimators
of the vertical actions are uncorrelated with $\Delta L_z$, and that
the mean and median changes are close to zero.  Comparison with the
systematic changes in Fig.~\ref{fig:run94ezchlch} suggest that $J_z$
is, in fact, conserved.  To see this, note that changes in $L_z$
shift the home radius of each particle to a region where the average
vertical restoring force differs, the amplitudes of the vertical
oscillations increase due to the weaker average restoring force when
$\Delta L_z > 0$, and conversely decrease for $\Delta L_z > 0$.  Thus
if an estimator of vertical motion is not conserved, we should expect
a systematic variation with $\Delta L_z$, as we observe for all the
energy estimators in Fig.~\ref{fig:run94ezchlch}.  The fact that there
is no systematic variation of the median or mean change in vertical
action suggest that it is conserved.

If $J_z$ is a conserved quantity, it may seem puzzling that the rms
changes in our measurements are not smaller.  Both $J_{z, \rm epi}$
and $J_z|_{R,\phi}$ are approximate, which could be responsible for
apparent substantial changes, but changes in the ``exact'' estimate of
$J_z$ remain significant, even for simulation T.  It is possible that
our estimate of $\Delta J_z$ is inaccurate.  For example, we must
eliminate non-axisymmetric structure from the potential to compute
$J_z$, introducing small errors when the potential is mildly
non-axisymmetric.  Larger differences could arise, however, if the
invariant curve changes its character, especially since entering or
leaving a trapped area in phase space is not an adiabatic change.  It
is possible an orbit that appears to be unaffected by resonant islands
in phase space at the initial and final times could have experienced
trapping about resonant islands for some of the intermediate
evolution, allowing $J_z$ to change even when changes to the potential
are indeed slow.

We have verified that $J_z$ is conserved to a part in $10^4$ in a
further simulation with a frozen, axisymmetric potential.  Note this
is not a totally trivial test, since we use the $N$-body integrator
and the grid-determined accelerations to advance the motion for $\sim
350$ dynamical times before making our second estimate.  Therefore the
non-zero changes in simulation T are indeed caused by changes to the
potential.  Even though the potential variations are small,
substantial action changes indicate that changes to the orbit were
non-adiabatic, which can happen for the reason given in the previous
paragraph.  Our ``initial'' $J_z$ values are estimated at
$t=64R_i/V_0$, when we believe the model has relaxed from the initial
set up; we therefore suspect that the small potential variations are
more probably driven by particle noise, which can be enhanced by
collective modes \citep{Wein98}.  This hypothesis is supported by our
finding variations in $J_z$ that were about ten times greater than in
simulation T when we employed 100 times fewer particles.

\section{Summary}
We have presented a quantitative study of the extent of radial
migration in both thin and thick discs in response to a single spiral
wave.  We find angular momentum changes in the thick disc are
generally smaller than those in the thin, although the tail to the
largest changes in each population is almost equally extensive.

We have introduced populations of test particles into a number of our
simulations in order to determine how changes in $L_z$ vary with disc
thickness and with radial velocity dispersion when subject to the same
spiral wave, finding an exponential decrease in $\rmsLz$, as shown in
Fig.~\ref{fig:run102103rmslch}.

We find that spirals of smaller spatial scale cause smaller changes.
When we were careful to change all the properties of the model by the
appropriate factors, we were able to account for smaller changes to
$\rmsLz$ as being due to a combination of the weakened spiral
amplitude near corotation and the change in the value of $m$.
Furthermore, we found evidence that the saturation amplitude scales
inversely as $m$, in line with the theory developed by \citet{S02}.
Note the simple scaling holds true only when the principal dynamical
properties, such as $Q$, $X$, thickness, and gravity softening, are
held fixed relative to the scale of the mode.  Nevertheless, it seems
reasonable to expect smaller changes to $\rmsLz$ in general for
spirals higher $m$.  The exponential decrease of $\rmsLz$ with
increasing disc thickness applies only for different populations
subject to the same spiral perturbation.

We have also run slightly more realistic simulations to follow the
extent of churning in both thick and thin discs that are subject to
large numbers of transient spiral waves having a variety of rotational
symmetries.  Fig.~\ref{fig:run104rhfrhi} shows that changes in the
home radii of thick disc particles are smaller on average than those
of the thin disc, but again the tails of the distributions in both
populations are almost co-extensive.  As found in previous work
\citep{F94, R98, G99, D06, M11, B11}, the formation of a bar also
causes substantial angular momentum changes within a disc, but we find
that the churning effect from multiple spiral patterns still dominates
changes in the outer disc after the bar has formed.

We find that vertical action is conserved during radial migration,
despite the fact that relative changes in our estimated $J_z$, which
can be measured for only about half the particles, are as large as
$\sim 15\%$.  Because the vertical restoring force to the mid-plane
decreases outwards, an increased (decreased) home radius causes the
particle to experience a systematically weaker (stronger) vertical
restoring force, making it impossible for both vertical energy and
action to be conserved.  We find a clear systematic variation of the
vertical energy with $\Delta L_z$, but none with $\Delta J_z$ leading
us to conclude that vertical action is the conserved quantity.  The
residual scatter in our measured values of $\Delta J_z$ could be
caused by trapping and escape from multiply-periodic resonant islands
in phase space, as well as numerical jitter in the $N$-body potential,
as we discuss at the end of \S6.  In the absence of these complications,
we believe that the vertical action is conserved.

Thus conservation of vertical action, and not of vertical energy,
should be used to prescribe the changes to vertical motion in models
of chemo-dynamic evolution.  It would be especially useful to obtain
diffusion coefficients that include the effects of radial migration as
functions of radius and height due to a transient spiral or a
combination of consecutive spirals with various corotation radii.  We
leave this work to a future paper.

\section*{Acknowledgment}
We thank the referee, Rok Ro\u skar, for a thorough and helpful
report.  This work was supported in part by NSF grant AST-1108977 to
JAS.


\appendix
\section[]{Numerical estimate of vertical action}

\begin{figure*}
\includegraphics[width=17cm]{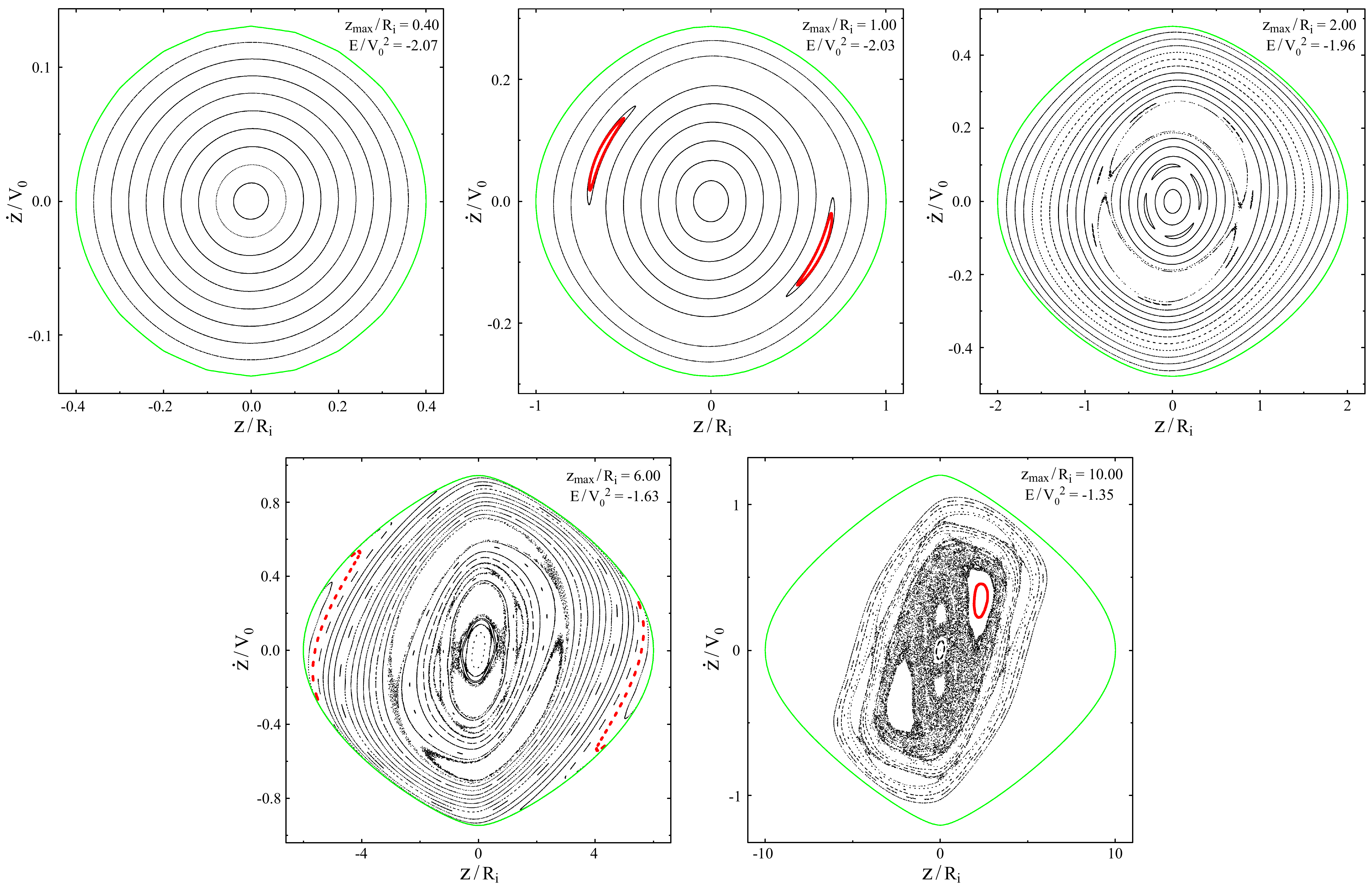}
\caption{Surfaces of section in the $(z,\dot z)$ plane, when $R=R_h$
  and $\dot R > 0V_0$, for test particles in the initial potential of
  simulation M2.  All particles in each panel have the same total
  energy and corresponding $z_{rm max}$, which are noted in the top right
  corner.  The outer most green curves mark the zero-velocity curves, at
  which $\dot R = 0V_0$ for the given $z$, $L_z$ and $E$.  Finally, the
  bold red curves are from the three orbits shown in
  Fig.~\ref{fig:run94vactorbts}.}
\label{fig:run94vactlecnst}
\end{figure*}

\begin{figure}
\includegraphics[width=84mm]{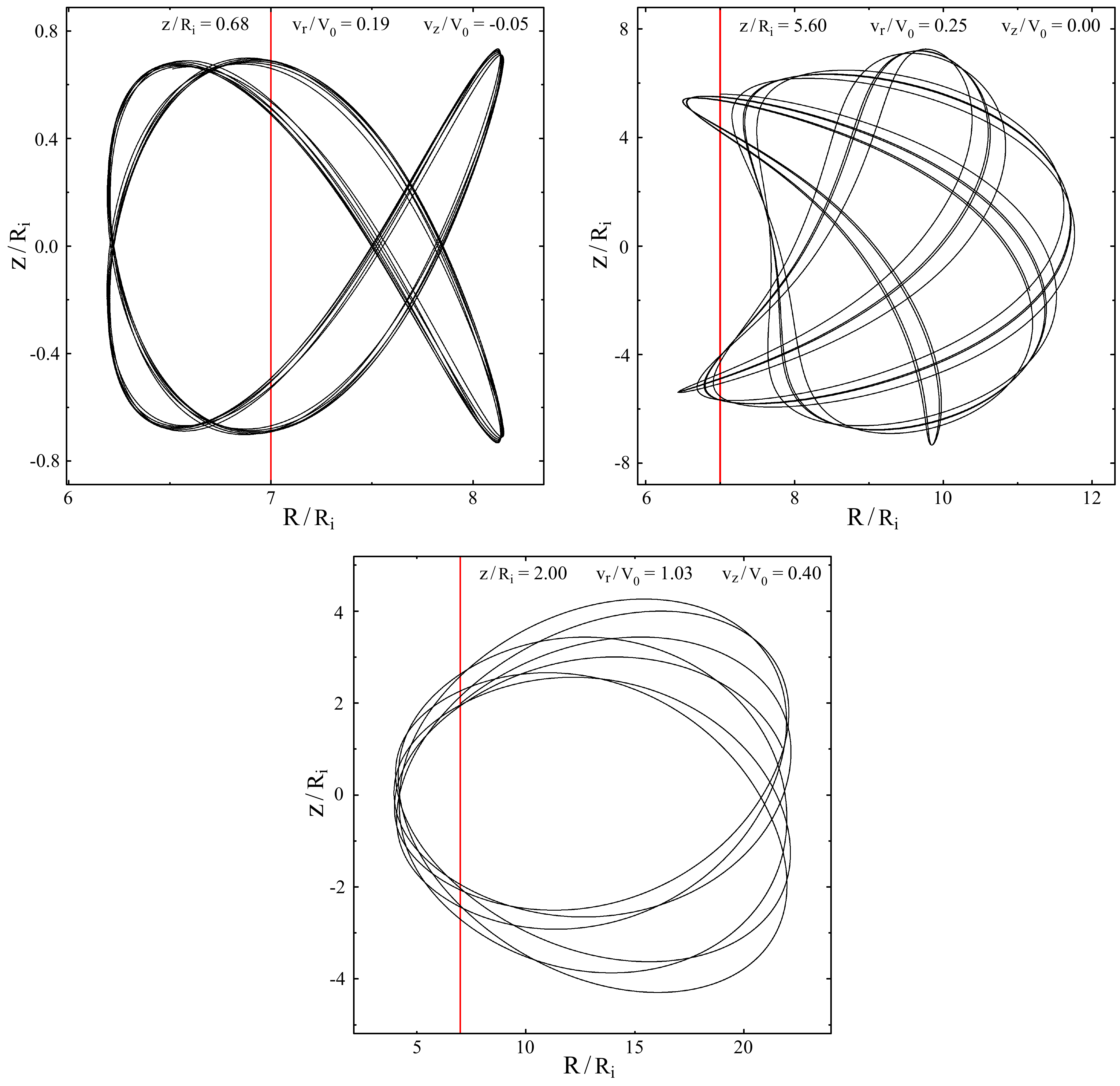}
\caption{Orbits in the meridional plane of the three test particles
  drawn in bold red in the second, fourth and fifth panels of
  Fig.~\ref{fig:run94vactlecnst}.  Note that we show only a small
  fraction of the full orbit used to produce the invariant curves in
  the $(z,\dot z)$ plane.  The vertical red lines mark $R_h$ at which
  the SoS is constructed.  The initial vertical position and radial
  and vertical velocities of each of these three particles are noted.}
\label{fig:run94vactorbts}
\end{figure}

Fig.~\ref{fig:run94vactlecnst} illustrates the SoS for five different
energies, all for $L_z=7$.  Each invariant curve is generated by a
test particle in the initial potential of M2.  All the particles in
one panel have the same energy and angular momentum (noted in the top
right corner), but differ in the extent of their vertical motion. 
The zero-velocity curve, where a particle's radial velocity must be
zero for the $L_z$ and $E$ values adopted is the outer most green curve.
We see that most of phase space is regular, but not all consequents
mark out simple closed invariant curves; some significant fraction of
phase space is affected by islands in the SoS that surround multiply
periodic orbits.  Chaotic motion, in which consequents fill an area
rather than lie on a closed curve in the SoS, is extensive only for
the greatest energy.

The paths in the meridional plane of three multiply periodic orbits
are shown in Fig.~\ref{fig:run94vactorbts}.  They correspond to the
bold red invariant curves in the second, fourth, and fifth SoS panels of
Fig.~\ref{fig:run94vactlecnst}.  The last orbit has almost equal
radial and vertical periods and circulates only clockwise in the
meridional plane.  Since we plot only when the particle passes $R_h$
with $\dot R > 0V_0$, the consequents lie only in the first quadrant of the
$(z,\dot z)$ plane.  Were we to plot the other crossing instead, where
$\dot R < 0V_0$, the invariant curve would be a reflected image in the
second quadrant.  Also, had we chosen the negative root for initial
radial velocity, the particle would have circulated only
counterclockwise and yielded a $180^\circ$-rotationally symmetric
invariant curve in the third quadrant.  This accounts for the skewness
of the invariant curves in the SoS, which diminishes for particles of
lower energy.

The area in the SoS enclosed by only those orbits having a simple
invariant curve yields a numerical estimate of the vertical action
$J_z$.\footnote{A regular orbit trapped about a resonant island has a
  different set of actions that are not of interest here.}  We find
all the consequents for a particle as its motion advances over
$7 \times 10^4$ dynamical times from its position in the frozen,
azimuthally averaged, potential, and make a numerical estimate of the
area integral (eq.~\ref{vertact}).

We repeat this exercise at both the initial and final times in the
simulation.

\label{lastpage}

\end{document}